\documentclass{article}

\usepackage[english]{babel}

\usepackage[letterpaper,top=2cm,bottom=2cm,left=3cm,right=3cm,marginparwidth=1.75cm]{geometry}

\usepackage{amsmath}
\usepackage{amsmath,amssymb,amsfonts}
\usepackage{graphicx}
\usepackage[colorlinks=true, allcolors=blue]{hyperref}
\usepackage{textcomp}
\usepackage{color,soul}
\usepackage{dblfloatfix}
\usepackage{multirow}
\usepackage[table,xcdraw]{xcolor}
\usepackage[listings]{tcolorbox}
\usepackage{pdflscape}
\usepackage{fontawesome5}
\usepackage{mdframed}
\usepackage{blindtext}

\newtcolorbox[list inside=mybox,auto counter,number within=section]{MyBox}{colbacktitle=yellow,coltitle=black,title={MyBox \thetcbcounter}}

\title{\textsf{A Reference Architecture for Quantum Computing as a Service}}
\author{$^1$Aakash Ahmad, $^2$Ahmed B. Altamimi, $^3$Jamal Aqib}
\date{%
    $^1$School of Computing and Communications. Lancaster University Leipzig, Germany\\ \url{a.ahmad13@lancaster.ac.uk}
    \\
    $^2$Department of Computer Engineering, University of Ha'il, Saudi Arabia\\
    \url{altamimi.a.uoh@gmail.com}
    \\
    $^3$Department of Information and Computer Science, University of Ha'il, Saudi Arabia\\
    \url{j.aqib@uoh.edu.sa}
    \\[2ex]%
}

\begin{document}
\maketitle

\begin{abstract}
Quantum computers (QCs) aim to disrupt the status-quo of computing – replacing traditional systems and platforms that are driven by digital circuits and modular software – with hardware and software that operates on the principle of quantum mechanics. QCs that rely on quantum mechanics can exploit quantum circuits (i.e., quantum bits for manipulating quantum gates) to achieve ‘quantum computational supremacy’ over traditional, i.e., digital computing systems. Despite being in a state of their infancy due to hardware limitations or lack of software ecosystem, QCs have started to demonstrate their data processing superiority in certain application areas including bio-inspired computing, cryptography, and tackling optimization problems. Currently, the issues that impede mass-scale adoption of quantum systems are rooted in the fact that building, maintaining, and/or programming QCs is a complex and radically distinct engineering paradigm when compared to challenges of classical computing and software engineering. Quantum service orientation is seen as a solution that synergises the research on service computing and quantum software engineering (QSE) to allow developers and users to build and utilise quantum software services based on pay-per-shot utility computing model. The pay-per-shot model represents a single execution of instruction on quantum processing unit and it allows vendors (e.g., Amazon Braket) to offer their QC platforms, simulators, software services etc. to enterprises and individuals who do not need to own or maintain quantum systems. Existing research lacks solutions in terms of empirically grounded processes, patterns, and guidelines to architect and implement quantum computing as a service. This research contributes by (i) developing a reference architecture for enabling quantum computing as a service, (ii) implementing microservices with the quantum-classic split pattern as an architectural use-case, and (iii) evaluating the reference architecture based on feedback by 22 practitioners. In the QSE context, the research focuses on unifying architectural methods and service-orientation patterns to promote reuse knowledge and best practices to tackle emerging and futuristic challenges of architecting and implementing Quantum Computing as a Service (QCaaS).
\end{abstract}


\section{Introduction}
\label{sec:introduction}

The emergence of quantum computers (QCs) has started to gradually disrupt traditional computing technologies - systems driven by binary logic and digital circuits - with machines that rely on quantum circuits to achieve computational efficiency \cite{1_Supremecy} \cite{2_QSE}. QCs exploit the fundamental of quantum mechanics via programmable Quantum Bits (QuBits) that operationalise Quantum Gates (QuGates) to execute some calculations exponentially faster than any `traditional computer' \cite{3_Qubits}. QCs represent a unification of hardware, i.e., quantum circuitry (QuBits mapped to QuGates) and software, i.e., quantum algorithms that manipulate the hardware, and network that can transmit and receive QuBit-encoded information \cite{4_QCSystem}. In quantum-era computing, QCs are undergoing a continuous evolution from their inception to a gradual maturity. However, such systems have been successful in mimicking biological systems and chemical reactions, solving optimization problems, and empowering fundamental science and existing technologies that are driven by quantum information processing \cite{5_QCPerformance}. To attain strategic benefits and developing commercial competencies associated with QCs, academic research \cite{1_Supremecy} \cite{2_QSE}, industrial projects (e.g., Amazon Braket \cite{6_AmazonBraket}), and technology funding consortiums (e.g., Quantum Flagship \cite{7_QunatumFlagship} are competing in a so-called race towards building quantum economies. Global policymakers and state representatives at the most recent World Economic Forum (WEF) advocated for building quantum economies that currently represent public and private investment worth \$35.5 billion \cite{8_WorldEconomicForum}. Despite the scientific benefits and commercial opportunities linked with QCs, a plethora of issues such as limited hardware, lack of software ecosystems, quantum noise, and scarcity of professional expertise in QSE domain hinders wide-scale adoption of quantum systems and technologies \cite{9_QuantumChallanges}. From users' perspective, enterprises and individuals lack access to quantum computing resources for tackling computationally hard tasks due to a multitude of challenges that may range from costs and economy of acquiring or maintaining QCs, quantum error rate, immature technology and/or lack of quantum software services \cite{10_QDivide}. From QSE perspective, engineers find themselves underprepared to tackle the complexities of quantum mechanics, handling QuBits/QuGates, knowledge of quantum programming, and workflows to develop software applications that can be executed on QC platforms \cite{15_QAAS, 16_QAASGateway}. This leads to a situation that is referred to as the quantum divide, representing a strategic and computational disparity between entities or states who have access and the ones that lack access to quantum systems and technologies \cite{10_QDivide}. To minimise this divide, initiatives across the world such as the Quantum Flagship \cite{7_QunatumFlagship} and National Quantum Initiative \cite{11_NationalQunatum} are focused on supporting the development of software ecosystems, networking technologies, and human expertise for the alleged quantum leap in computing \cite{2_QSE}.

\vspace{0.3 em}
\textbf{Service-orientation for QCs:} Service-oriented systems are viewed as the enablers of utility computing model, allowing pay-per-usage of software applications and hardware resources that are made available as a computing services, to be used by individuals and enterprises \cite{12_ServiceComputing}. The providers of utility computing (i.e., service vendors)  offer a variety of  services to their customers (i.e., service users) that range from data storage, video streaming, and entertainment, to resource-sharing applications, representing a multi-billion dollar industry in service economies \cite{13_SCRevenue}. Central to the success of service-orientation is the concept of as-a-Service (aaS) or anything-as-a-Service (*aaS) model that provides any-time, any-where distributed access to \textit{infrastructures} (e.g., Microsoft Azure), \textit{platforms} (e.g., Google App Engine) and \textit{software} (e.g., Cisco WebEx) as a service \cite{14_AAS}. Attuned to the practices of service-orientation is the concept of quantum service computing that allows quantum vendors to provide and quantum users to request quantum hardware and software resources available via network \cite{15_QAAS} To breach the quantum divide \cite{10_QDivide}, quantum service-orientation can enable service requesters (QC users) to access resources offered by service providers (QC vendors). Figure \ref{fig1:intro} conceptualises a quantum service computing model where the users can utilise a multitude of resources such as quantum processing units, simulators, storage, algorithms, and software applications to tackle computationally challenging tasks via QCs. On the flip side, QC vendors see quantum service computing as an opportunistic business model to generate revenue streams by offering pay-per-shot resources or crowd-sourced testing of their under-developed quantum platforms \cite{6_AmazonBraket, 16_QAASGateway}. In QC context, a shot represents a single execution of quantum instruction on a quantum processor unit. Considering quantum service-orientation, as envisioned in Figure \ref{fig1:intro}, there is a need for research and development, rooted in empirically grounded processes, patterns, and tools to architect and implement Quantum Computing as a Service (QCaaS).

\vspace{0.3 em}

\textbf{Research context and objectives:} Quantum software engineering is regarded as a recently emerged genre of software engineering (SE) that aims to apply the principles, processes, and practices of SE to systemise the development of quantum software systems and services \cite{2_QSE} \cite{17_QSE}. QSE empowers the role of software designers and developers who can exploit processes that support structured development, architectural models for design visualisation, and patterns as best practices to engineer software that can be executed on QCs \cite{18_QSA}. QSE approaches can help developers to abstract the complexities of quantum mechanics such as mapping operations of QuBits and QuGates to software components that can be transformed to modules of quantum source code via model-driven engineering \cite{19_MDQSE} The objective of this research is \textit{`to support pattern-based architecting - enabling reuse of best practices and exploiting architectural modeling - to engineer software services in the context of QSE and for enabling QCaaS'}.  By relying on QSE, we adopted a stepwise approach to architect and implement (proof-of-the-concept) QCaaS for quantum information processing. The proposed research aims to synergise the principles of QSE and practices of service-oriented computing to (i) develop a reference architecture that acts as a blue-print for (ii) implementing a prototype of quantum computing as a service. We engaged a total of 22 QSE professionals, from 12 countries across 6 continents, experienced in working with various QC platforms (e.g., Amazon Braket, Quantum Azure) to evaluate the suitability and usability of the reference architecture \cite{20_ISOQuality}. The reference architecture can enable engineers to abstract complexities of quantum source code into architectural components, apply reuse knowledge via quantum software patterns, and adopt best practices such as microservices architecture style to develop QCaaS solutions \cite{16_QAASGateway}. We developed a QCaaS solution by using microservices architecture, applied service orchestration and quantum-classic split patterns, and executed quantum service-orientation on the Microsoft Azure platform. Primary contributions of this research include:

\begin{figure*}
\centering
\includegraphics[width=0.6\textwidth]{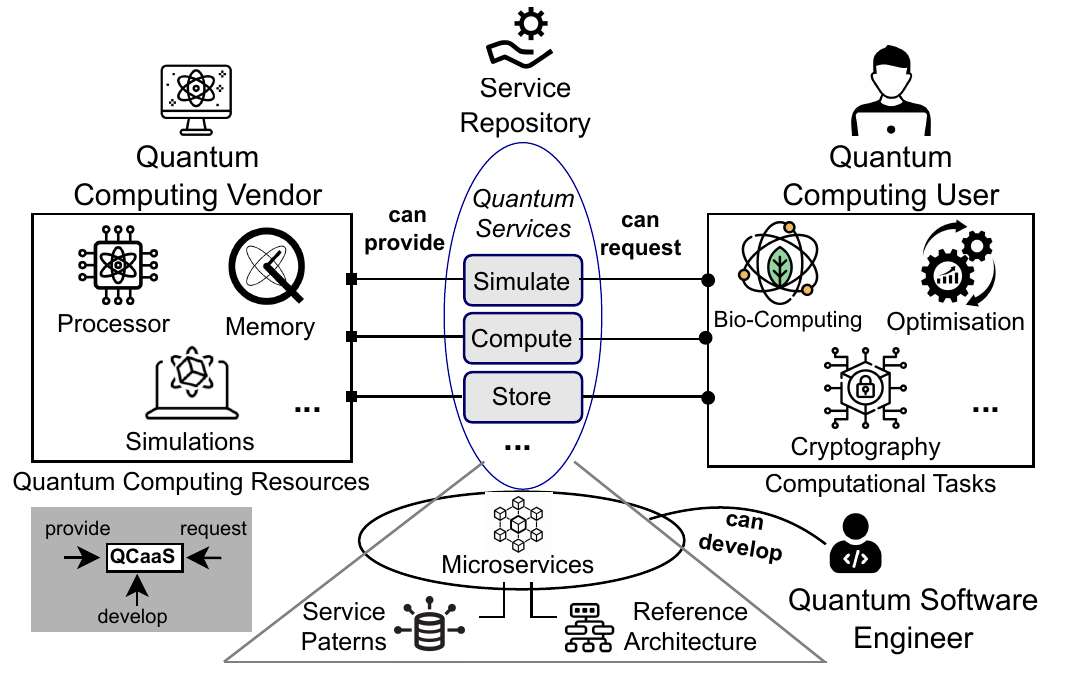}
\caption{\label{fig1:intro} An Overview of Quantum Service Computing.}
\end{figure*}

\begin{itemize}
\item An empirically grounded reference architecture, rooted in the systematic mapping of published research \cite{15_QAAS} and IBM service development lifecycle \cite{IBMSOA}, to provide a blue-print and point of reference to architect software-intensive systems and services for QCaaS. 

\item A proof-of-the-concept that demonstrates architecture-centric and pattern-driven implementation of quantum software services that can be executed on a quantum computing platform.

\item Practitioners’ evaluation of the reference architecture that provides recommendations and guidelines to design and develop solutions for QCaaS.
\end{itemize}

The study represents a pioneering effort in architecting QCaaS solutions and it requires empiricism, diverse usecase, and further experimentation as part of future research. This study can help academic researchers to understand the role of reference architectures, patterns, micro-servicing etc. and help them formulate new hypotheses for investigating emerging and futuristic challenges of QSE in the context of QCaaS. Practitioners can explore and extend the reference architecture (system blueprint) and reuse knowledge (patterns) that can be adopted to develop solutions for QCaaS.

Rest of this paper is organised as follows. Section \ref{sec:Background} discusses background details on quantum systems and quantum service computing. Section \ref{sec:ResearchMethod} presents the research method to conduct this study. Section \ref{sec:RefArch} details the design and interpretation of the reference architecture for quantum computing as a service. Section \ref{sec:implement} presents a proof-of-the-concept implementation of the reference architecture. Section \ref{sec:ArchEvaluation} discusses evaluation of the reference architecture. Section \ref{sec:RelatedWork} presents the related work to rationalise the scope and contributions of the proposed research. Section \ref{sec:Threats} details threats to the validity of this research. Section \ref{sec:Conclusions} presents conclusions and dimensions of future work.
\section{Context: Service Orientation for Quantum Computing}
\label{sec:Background}
\begin{figure*}
\centering
\includegraphics[width=0.80\textwidth]{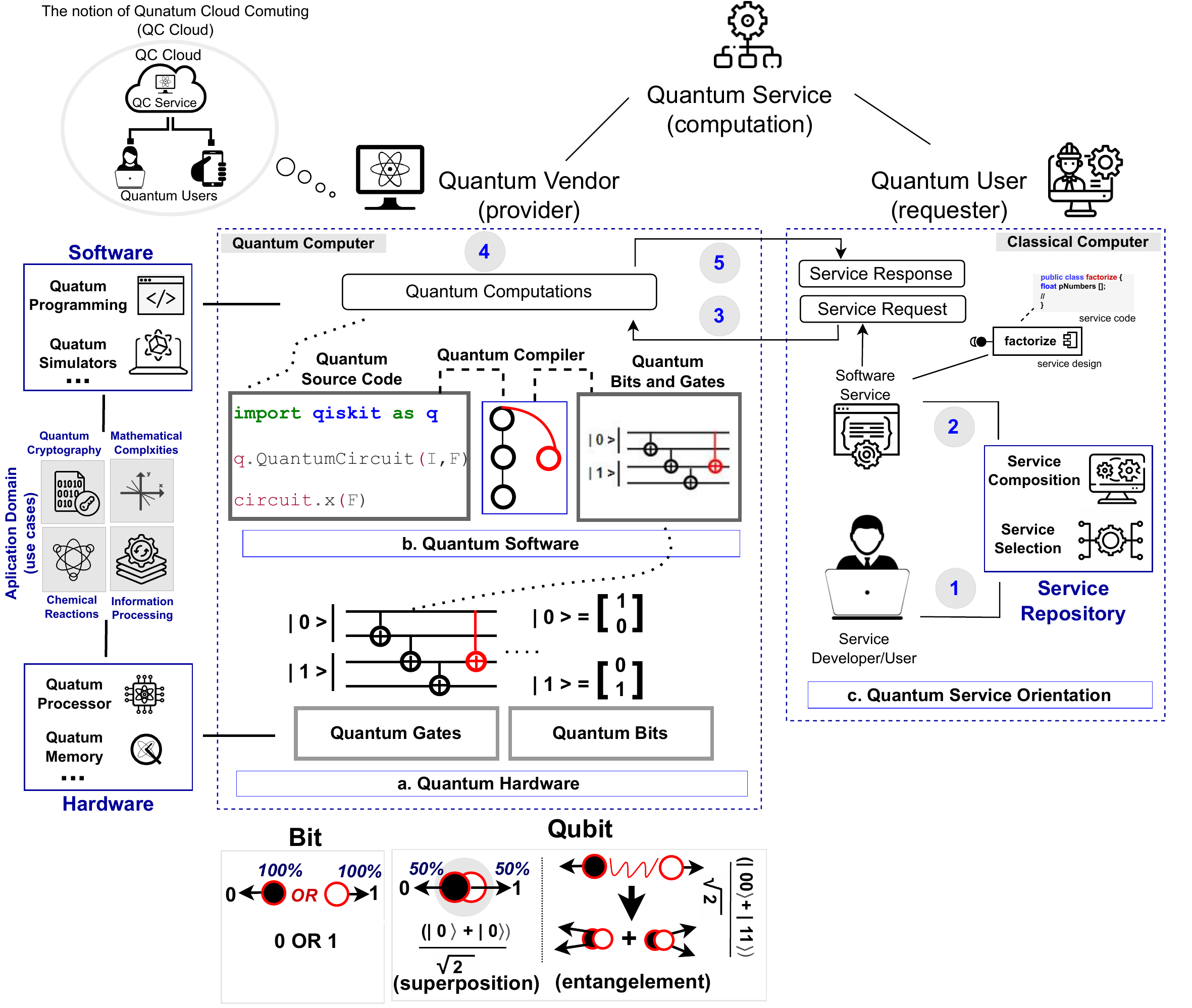}
\caption{\label{fig2:qsc} A General View of the Quantum Service Computing Systems.}
\end{figure*}

This section presents background on quantum computing (Section \ref{sec:2_1}) and service-orientation for quantum systems (Section \ref{sec:2_2}) to contextualise service-orientation for quantum computing. We use the illustrations in Figure \ref{fig2:qsc} that shows building blocks and conceptualisation of quantum computing as a service. The concepts and terminologies introduced in this section are used throughout the paper.

\subsection{Quantum Computing Systems}\label{sec:2_1}
To gain strategic advantages in the quantum race, i.e., attaining technical and commercial benefits of quantum computing, research and development efforts are focused on engineering both the hardware and software systems that can be unified into a practical quantum computer \cite{4_QCSystem} \cite{1_Supremecy}. Application domains or practical use cases of QC systems such as quantum cryptography or bio-inspired computing can exploit quantum hardware resources by means of quantum software systems and applications that can manipulate the hardware \cite{2_QSE}. We discuss the quantum computing system from both the hardware and software perspective as in Figure \ref{fig2:qsc} a). Fundamental to achieving quantum computations are Quantum Bits (QuBits) that represent the basic unit of quantum information processing by manipulating Quantum Gates (QuGates) \cite{3_Qubits}. Traditional Binary Digits (Bits) in classical systems (i.e., digital computers) are represented as [1, 0] where 1 represents the computation state as ON and 0 represents the state as OFF to manipulate binary gates in digital circuits.  In comparison, a QuBit represents a two-state quantum computer expressed as $|0\rangle$ and $|1\rangle$. Specifically, the state of a single QuBit can be expressed as $|0\rangle = \begin{bmatrix} 1 \\ 0 \end{bmatrix}$ and $|1\rangle = \begin{bmatrix} 0 \\ 1 \end{bmatrix}$. When compared to a Bit, quantum superposition allows a QuBit to attain a liner combination of both states:

\begin{equation}\label{EQ-1}
 |0\rangle  =  \left[ \begin{array}{c} 1 \\ 0 \end{array} \right] ~~~~~ + ~~~~~  |1\rangle  =  \left[ \begin{array}{c} 0 \\ 1 \end{array} \right]   
\end{equation}

In Figure \ref{fig2:qsc} a), we distinguish between a Bit and Qubit. Unlike the Bit that can manipulate an electronic circuit as On and Off, a Qubit uses the properties of quantum mechanics to be a 1 or 0 or attain any value in between. Specifically, a Bit can take a value of `0' or `1' representing as either `Off:0' or `On:1' with 100\% probability (left, Figure \ref{fig2:qsc}). In comparison, a Qubit can be in a state of $|0\rangle$ or $|1\rangle$ or in a superposition state with 50\% $|0\rangle$ and 50\%  $|1\rangle$ (middle). In addition, two Qubits can be entangled (right), and entangled qubits are linked in a way that observing (i.e., measuring) one of the QuBit, can reveal the state of other QuBit. There is an abundance of literature and use cases that discuss theoretical aspects of quantum physics and its application to quantum systems, extended details of QuBits and QuGates to develop and operate the QC systems are reported in studies like \cite{21_QuBitProgram} and \cite{3_Qubits}. To utilise the QC resources such as quantum processor and memory, there is a need for control software that can program QuBits to manage QuGates of a QC system. Quantum software systems rely on quantum source code compilers that allow quantum algorithm designers and code developers to write, build, and execute software for quantum computers \cite{22_QPL}. For example, a programmer can use a quantum programming language such as Q\# (by Microsoft) or Qiskit (by Google) to use quantum compilers for enabling programable quantum computations \cite{23_QCCompiler}. By developing software systems that can manage and control quantum hardware, a number of applications such as quantum cryptography, bio-inspired computing, and quantum information processing can benefit from quantum supremacy in computing.

\subsection{Service-Orientation for Quantum Computing}\label{sec:2_2}
Service computing follows the Service Oriented Architecture (SOA) style that allows service users to discover and utilise a multitude of available software services that encapsulate computing resources and applications offered by service providers \cite{12_ServiceComputing}. Figure \ref{fig2:qsc} b) shows SOA-styled quantum servicing where a QC user (i.e., service requester) can utilise the QC resources offered by quantum vendors (i.e., service provider) by means of quantum services \cite{15_QAAS}. Despite the computational superiority of QC systems in tackling certain classes of complex problems, when compared to traditional computers, quantum systems are in a state of infancy due to continuous evolution of hardware, immature technology, and limitations of quantum algorithms. In most cases, QC systems of today are not capable of executing quantum algorithms wrapped with a large amount of data, inputs, and outputs \cite{9_QuantumChallanges}. More specifically, contextualising the computation illustrations in Figure \ref{fig2:qsc}, large volumes of data in quantum algorithms require more QuBits and complex QuGates that result in deep quantum circuiting and consequently increased errors referred to as noisy intermediate-scale quantum (NISQ) \cite{24_NISQ}. To address issues like NISQ and to make quantum computers more practical in handling a significant amount of processing, the classic-quantum split pattern splits quantum software (having algorithms and data) into a classical part and a quantum part \cite{25_QPattern}. In principle, the classic-quantum split slices the overall quantum software or application into classical modules (pre/post-processing) and quantum modules (quantum computation) that result in hybrid applications \cite{25_QPattern}. One of the prime examples of the classic-quantum split patterns is Shor’s algorithm which involves quantum computations for finding the prime factors of an integer with its application in computer security and cryptography \cite{26_Shor}. The algorithm is composed of two parts. The first part of the algorithm turns the factoring problem into the problem of finding the period of a function and may be implemented classically. The second part finds the period using the quantum fourier transformation and is responsible for the quantum speedup. The quantum service can create (1) factors of numbers and provide this an input, (2) the factorization is done on a quantum computer and (3) results are returned for further processing on a local (classical) computer \cite{25_QPattern}.

Quantum computing vendors such as Amazon, IBM, and Google have started to offer their QC systems and infrastructures that can be utilised by individuals and organisations by means of quantum service computing \cite{29_QCCloud}. For example, Amazon Braket as a QC platform relies on Amazon Web Services (AWS) to support the execution of quantum applications \cite{6_AmazonBraket}. In addition to industrial development and commercialisation, academic research is also striving for solutions that can offer algorithms, hardware, and mathematical problems as services \cite{15_QAAS}. To harness QC as utility computing, there is a need to tailor existing principles and methods of service-orientation or develop new architectures and frameworks, empirically grounded processes to synergise QC and SOA research in the context QCaaS. The QCaaS also provides foundations for building a quantum computing cloud (QC cloud), led by technology giants like IBM to offer computation resources to the end-users based on cloud computing model \cite{30_IBMCloud}.

\section{Research Method}
\label{sec:ResearchMethod}
This section details the research method to conduct this study as illustrated in Figure \ref{fig3:method}. Each of the four steps of the research method, as visualised in Figure \ref{fig3:method}, are elaborated below in Section \ref{sec:3_1} - Section \ref{sec:3_4}.

\subsection{Step I - Conducting the Mapping Study}\label{sec:3_1}
\begin{figure}
\centering
\includegraphics[width=1.05\textwidth]{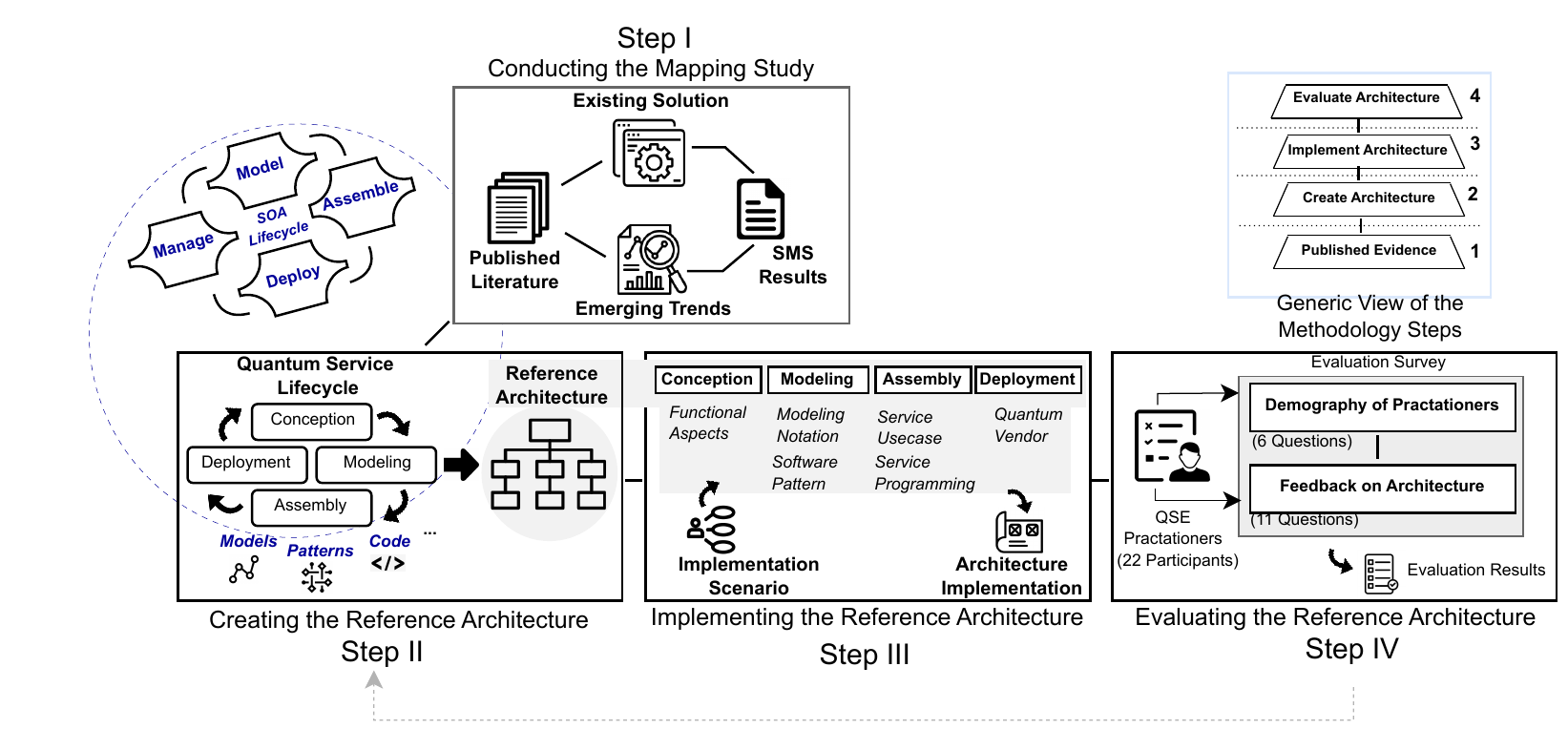}
\caption{\label{fig3:method}. Process and Steps of the Research Method}
\end{figure}

As per Figure \ref{fig3:method}, the initial step of the research method focused on conducting a systematic mapping study to analyse the existing literature, emerging trends, and prominent challenges of architecting and implementing quantum computing as a service. Systematic mapping study (SMS or mapping study for short) as an approach is grounded in evidence-based software engineering method for reliable and replicable identification, analysis and synthesis of data or facts on a topic under investigation. We followed the guidelines in \cite{27_Mapping} to conduct SMS of existing published research that enables or enhances architecting and development of quantum computing as a service. The results of SMS on QCaaS along with the process to conduct and document the study are detailed in \cite{15_QAAS}. We only highlight the core findings of the SMS in Table \ref{tab:SMSResults} that synopsises the available evidence, derived from published research and proposed solutions, to provide the basis for creating reference architecture for QCaaS (in Section \ref{sec:3_2}). A synoptic view of the SMS results, reflected in Table \ref{tab:SMSResults}, helped us identify four phases and six activities to support the conception, modelling, assembly, and deployment of QCaaS.   

\subsection{Step II - Creating the Reference Architecture}\label{sec:3_2}
From software engineering and architecting perspective, a reference architecture denotes a blue-print as a collection of documents, notations, or design artifacts etc. to provide a recommended structure to instantiate solutions in terms of software-intensive systems, services, and applications. Empiricism is central to design, develop, and validate the reference architectures. One specific example of an empirically grounded reference architecture derived from five industrial projects is presented in \cite{28_ArchModel} which suggests three main phases in terms of architectural analysis, architectural synthesis, and architectural evaluation to develop software systems. In the context of quantum software engineering, the research in \cite{18_QSA} extends the generic reference architecture from \cite{28_ArchModel} to propose a process and reference architecture for quantum software. Based on the reference architecture for general purpose software \cite{28_ArchModel} and architecture process for quantum software applications \cite{18_QSA}, we developed a reference architecture based on four phases from SOA lifecycle \cite{IBMSOA}.  Table \ref{tab:SMSResults} acts as a structured catalogue to organise four phases and six activities for architecting QCaaS. 

\begin{landscape}
\begin{table*}[]
\caption{Summary of the Core Findings form SMS \cite{15_QAAS}}
\begin{centering}
{\scriptsize
\begin{tabular}{|c|c|cc|cc|c|}
\hline
\multicolumn{7}{|c|}{\includegraphics[scale=0.85]{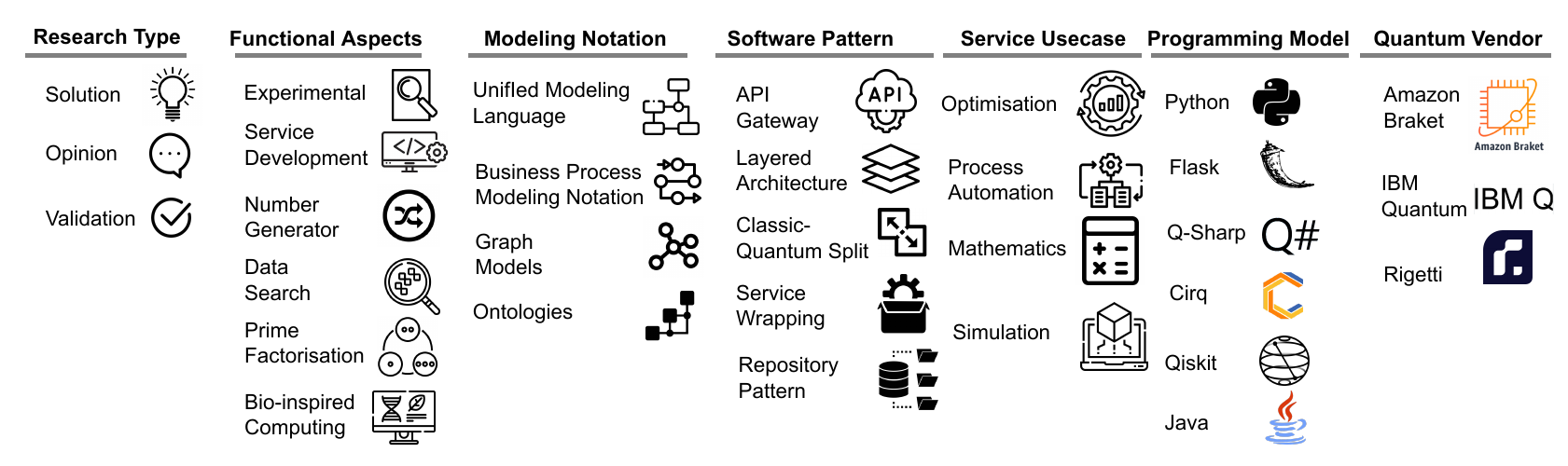}} \\ \hline
\multicolumn{1}{|l|}{} & \multicolumn{1}{c|}{\cellcolor[HTML]{BDD6EE}\textbf{Conception}} & \multicolumn{2}{c|}{\cellcolor[HTML]{BDD6EE}\textbf{Modeling}} & \multicolumn{2}{c|}{\cellcolor[HTML]{BDD6EE}\textbf{Assembly}} & \multicolumn{1}{c|}{\cellcolor[HTML]{BDD6EE}\textbf{Deployment}} \\ \cline{2-7}
\multirow{-2}{*}{\textbf{\begin{tabular}[c]{@{}c@{}}Study ID \\ \& \\ Research Type\end{tabular}}} &
  \cellcolor[HTML]{F2F2F2}\begin{tabular}[c]{@{}c@{}}Functional \\ Aspects\end{tabular} &
  \multicolumn{1}{c|}{\cellcolor[HTML]{F2F2F2}\begin{tabular}[c]{@{}c@{}}Modelling \\ Notation\end{tabular}} &
  \cellcolor[HTML]{F2F2F2}\begin{tabular}[c]{@{}c@{}}Software\\ Pattern\end{tabular} &
  \multicolumn{1}{c|}{\cellcolor[HTML]{F2F2F2}\begin{tabular}[c]{@{}c@{}}Service \\ Use case\end{tabular}} &
  \cellcolor[HTML]{F2F2F2}\begin{tabular}[c]{@{}c@{}}Service\\ Programming\end{tabular} &
  \cellcolor[HTML]{F2F2F2}\begin{tabular}[c]{@{}c@{}}Quantum\\ Platform/\\ Vendor\end{tabular} \\ \hline
\rowcolor[HTML]{FFFFFF}

\begin{tabular}[c]{@{}c@{}}{\includegraphics[scale=0.11]{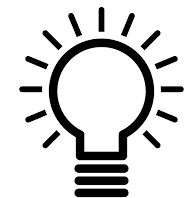}} \\ {[}S1{]}\end{tabular} &
  \begin{tabular}[c]{@{}c@{}}{\includegraphics[scale=0.04]{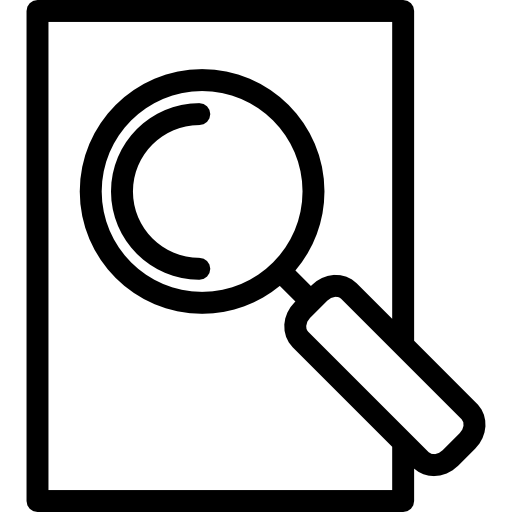}}\\    Quantum   Service \\ Delivery\end{tabular} &
  \multicolumn{1}{c|}{\cellcolor[HTML]{FFFFFF}\begin{tabular}[c]{@{}c@{}}{\includegraphics[scale=0.04]{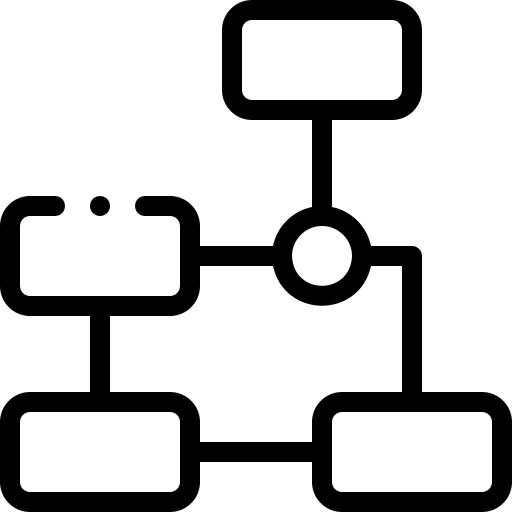}}   \\ Deployment\\ Diagram\end{tabular}} &
  \begin{tabular}[c]{@{}c@{}}{\includegraphics[scale=0.04]{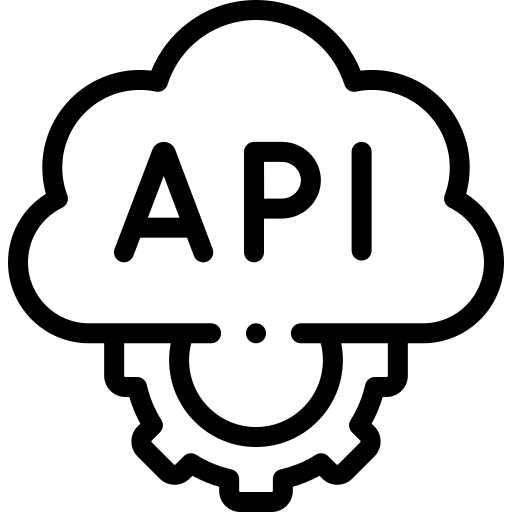}}\\ API Gateway\end{tabular} &
  \multicolumn{1}{c|}{\cellcolor[HTML]{FFFFFF}\begin{tabular}[c]{@{}c@{}}{\includegraphics[scale=0.04]{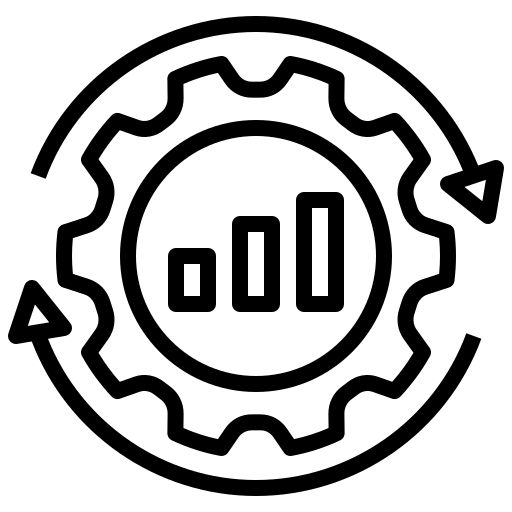}} \\ Optimal Service \\ Provider\end{tabular}} &

  \begin{tabular}[c]{@{}c@{}}{\includegraphics[scale=0.15]{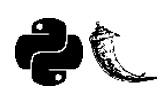}} \\ Python, Flask \end{tabular} &
  
  \begin{tabular}[c]{@{}c@{}}{\includegraphics[scale=0.20]{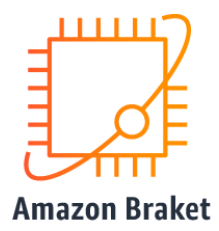}}\end{tabular} \\ \hline
\rowcolor[HTML]{FFFFFF} 
\begin{tabular}[c]{@{}c@{}}{\includegraphics[scale=0.04]{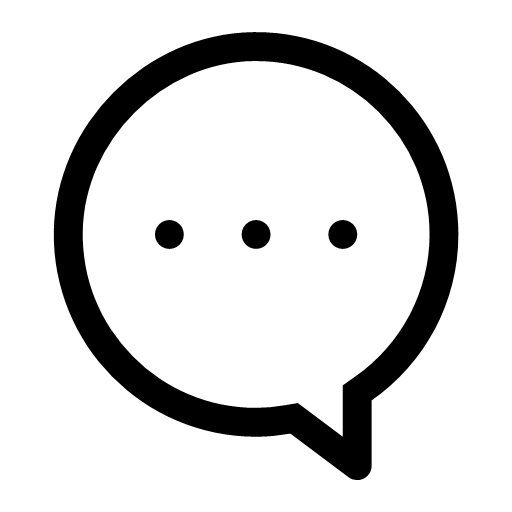}}\\   {[}S2{]}\end{tabular} &
  \begin{tabular}[c]{@{}c@{}}{\includegraphics[scale=0.07]{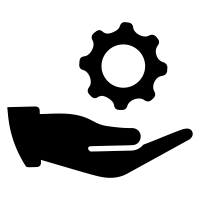}}   \\ Enterprise  Services \\ Development\end{tabular} &
  \multicolumn{1}{c|}{\cellcolor[HTML]{FFFFFF}\begin{tabular}[c]{@{}c@{}}{\includegraphics[scale=0.04]{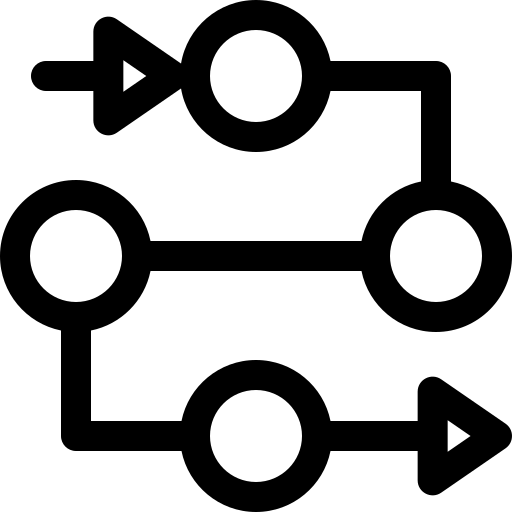}}\\   Business Process\end{tabular}} &

  \begin{tabular}[c]{@{}c@{}}{\includegraphics[scale=0.07]{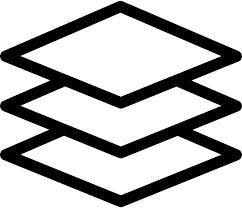}}  \\ Layered\\ Architecture\end{tabular} &
  
  \multicolumn{1}{c|}{\cellcolor[HTML]{FFFFFF}\begin{tabular}[c]{@{}c@{}}{\includegraphics[scale=0.04]{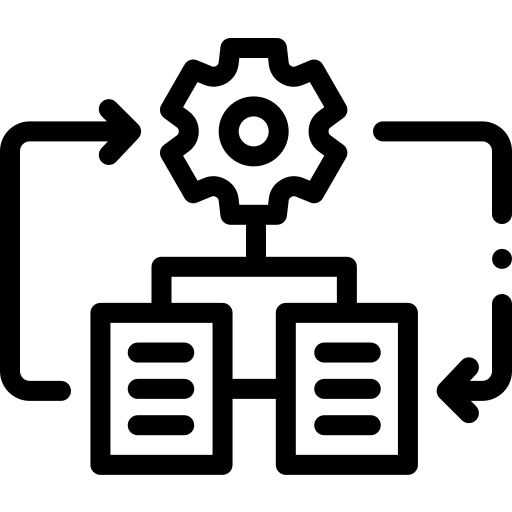}}   \\ Process Automation\end{tabular}} &
  
  \begin{tabular}[c]{@{}c@{}}{\includegraphics[scale=0.07]{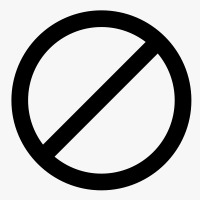}}\\No Evidence \end{tabular} &
  \begin{tabular}[c]{@{}c@{}}{\includegraphics[scale=0.07]{Figs/icons/Noevidence.png}}\\No Evidence \end{tabular} \\ \hline
\rowcolor[HTML]{FFFFFF} 
\begin{tabular}[c]{@{}c@{}}{\includegraphics[scale=0.11]{Figs/icons/1.jpg}}\\   {[}S3{]}\end{tabular} &
  \begin{tabular}[c]{@{}c@{}}{\includegraphics[scale=0.11]{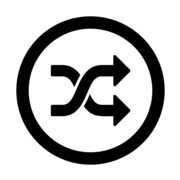}}   \\ Quantum Random \\ Number Generation\\ Quantum Search Algo\end{tabular} &
  \multicolumn{1}{c|}{\cellcolor[HTML]{FFFFFF}\begin{tabular}[c]{@{}c@{}}{\includegraphics[scale=0.04]{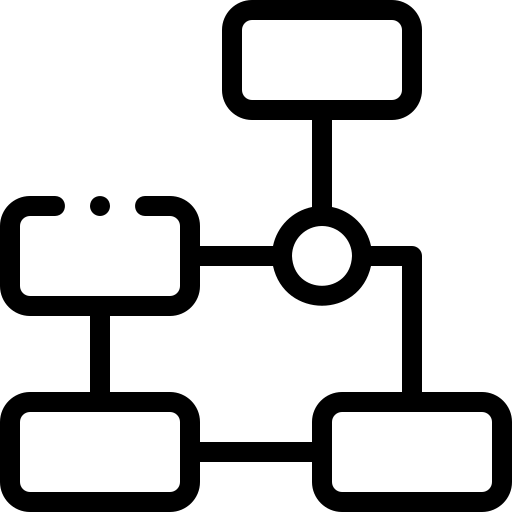}}   \\ Class, Sequence \\ Diagram\end{tabular}} &

  \begin{tabular}[c]{@{}c@{}}{\includegraphics[scale=0.07]{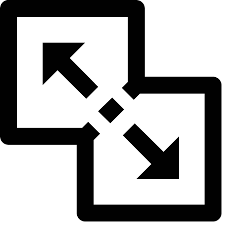}}\\Classic-
Quantum Split\end{tabular} &
  
  \multicolumn{1}{c|}{\cellcolor[HTML]{FFFFFF}\begin{tabular}[c]{@{}c@{}}{\includegraphics[scale=0.07]{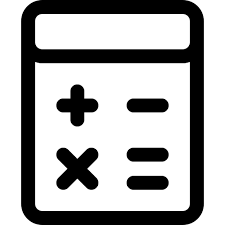}} \\ Mathematics\end{tabular}} &
  
  \begin{tabular}[c]{@{}c@{}}{\includegraphics[scale=0.11]{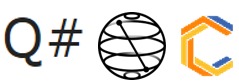}} \\Q sharp\end{tabular} &

  \begin{tabular}[c]{@{}c@{}}{\includegraphics[scale=0.11]{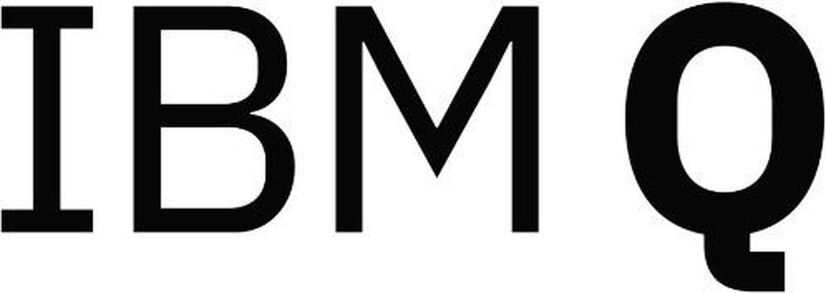}} \\IBM Quantum\end{tabular} \\ \hline
\rowcolor[HTML]{FFFFFF} 
\begin{tabular}[c]{@{}c@{}}{\includegraphics[scale=0.04]{Figs/icons/opinion.png}} \\ {[}S4{]}\end{tabular} &
  
  \begin{tabular}[c]{@{}c@{}}{\includegraphics[scale=0.11]{Figs/icons/numb.png}}\\ Integer Factorisation\end{tabular} &
  
  \multicolumn{1}{c|}{\cellcolor[HTML]{FFFFFF}\begin{tabular}[c]{@{}c@{}}{\includegraphics[scale=0.05]{Figs/icons/UMLS.png}}\\ Deployment Diagram\end{tabular}} &
  \begin{tabular}[c]{@{}c@{}}{\includegraphics[scale=0.11]{Figs/icons/1.jpg}} \\ Solution\end{tabular} &
  
  \multicolumn{1}{c|}{\cellcolor[HTML]{FFFFFF}\begin{tabular}[c]{@{}c@{}}{\includegraphics[scale=0.07]{Figs/icons/maths.png}} \\ Mathematics\end{tabular}} &
  
  \begin{tabular}[c]{@{}c@{}}{\includegraphics[scale=0.005]{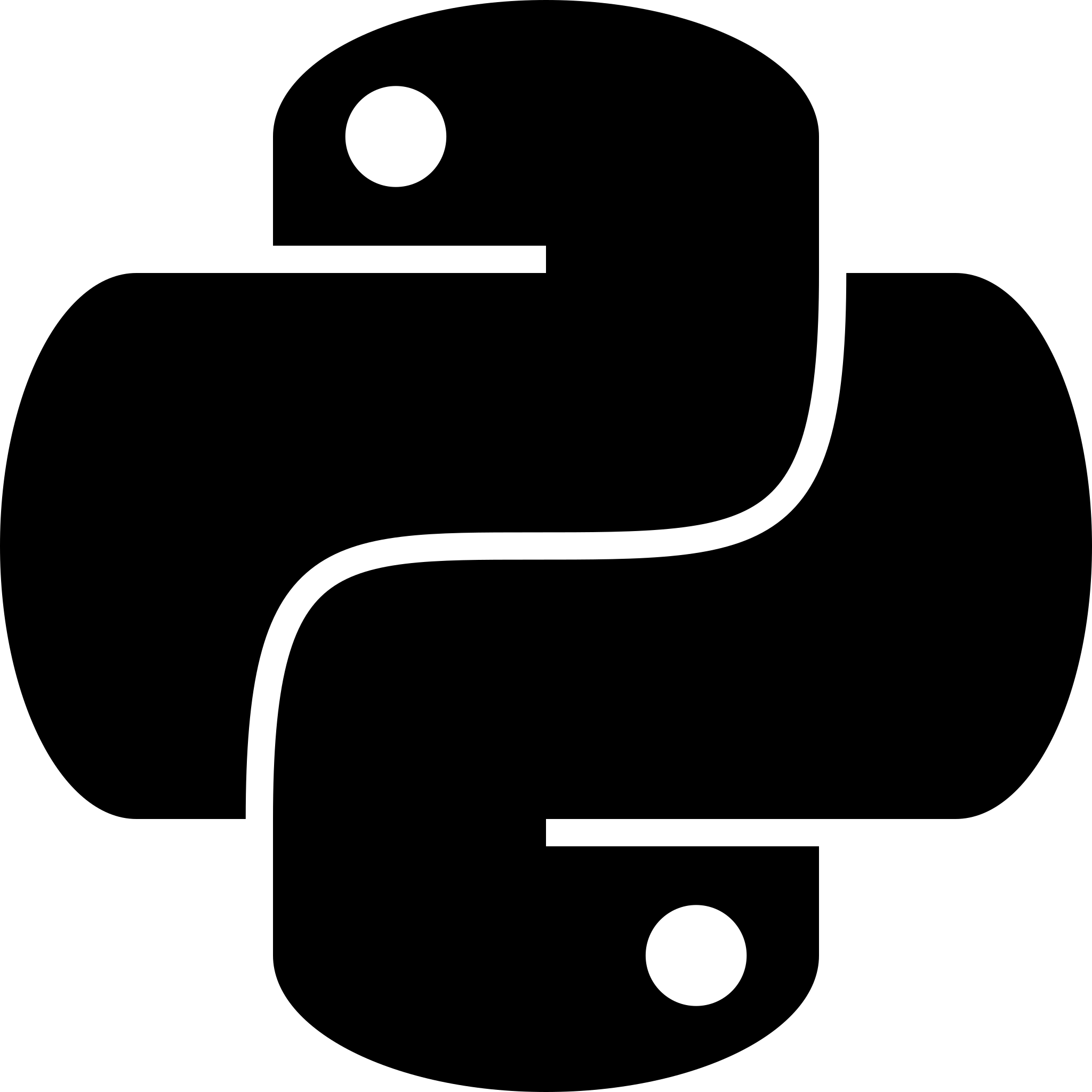}}\\Python\end{tabular} &
  
  \begin{tabular}[c]{@{}c@{}}{\includegraphics[scale=0.20]{Figs/icons/amazon.png}}\end{tabular} \\ \hline
  
  \rowcolor[HTML]{FFFFFF} 
\begin{tabular}[c]{@{}c@{}}{\includegraphics[scale=0.01]{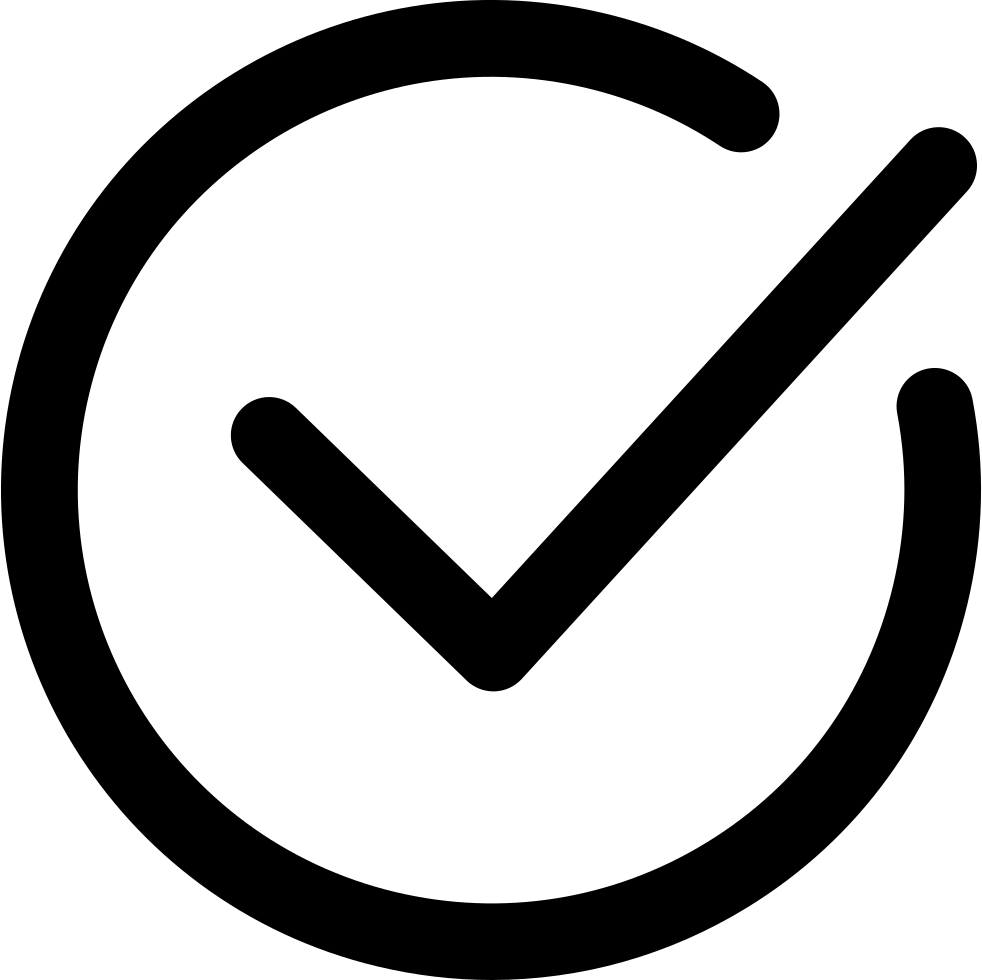}}  \\ {[}S5{]}\end{tabular} &
  \begin{tabular}[c]{@{}c@{}}{\includegraphics[scale=0.07]{Figs/icons/ServiceDelivery.png}}\\ Experimental Quantum\\   Service Computing)\end{tabular} &
  
  \multicolumn{1}{c|}{\cellcolor[HTML]{FFFFFF}\begin{tabular}[c]{@{}c@{}}{\includegraphics[scale=0.07]{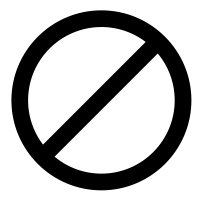}}\\No Evidence\end{tabular}} &
  
  \begin{tabular}[c]{@{}c@{}}{\includegraphics[scale=0.09]{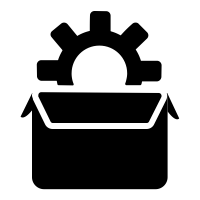}}\\Service  
Wrapping \end{tabular} &
  
  \multicolumn{1}{c|}{\cellcolor[HTML]{FFFFFF}\begin{tabular}[c]{@{}c@{}}{\includegraphics[scale=0.04]{Figs/icons/oPTIMISATIOPN.png}}\\Optimisation\end{tabular}} &

  \begin{tabular}[c]{@{}c@{}}{\includegraphics[scale=0.15]{Figs/icons/flask1.png}}\\Python, Flask\end{tabular} &
  
  \begin{tabular}[c]{@{}c@{}}{\includegraphics[scale=0.20]{Figs/icons/amazon.png}}\end{tabular} \\ \hline

\rowcolor[HTML]{FFFFFF} 
\begin{tabular}[c]{@{}c@{}}{\includegraphics[scale=0.11]{Figs/icons/1.jpg}}\\   {[}S6{]}\end{tabular} &
  \begin{tabular}[c]{@{}c@{}}{\includegraphics[scale=0.07]{Figs/icons/ServiceDelivery.png}}\\ Experimental Services\\ Algorithm\end{tabular} &
  \multicolumn{1}{c|}{\cellcolor[HTML]{FFFFFF}\begin{tabular}[c]{@{}c@{}}{\includegraphics[scale=0.04]{Figs/icons/UML.png}}\\ Sequence Diagrams\end{tabular}} &
  
  \begin{tabular}[c]{@{}c@{}}{\includegraphics[scale=0.04]{Figs/icons/API.png}}\\ API Gateway\end{tabular} &

  \multicolumn{1}{c|}{\cellcolor[HTML]{FFFFFF}\begin{tabular}[c]{@{}c@{}}{\includegraphics[scale=0.07]{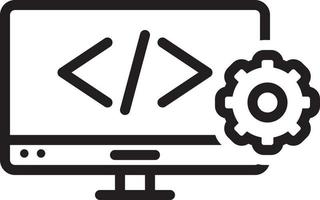}}\\ Algorithm as \\ a Service\end{tabular}} &
 
  \begin{tabular}[c]{@{}c@{}}{\includegraphics[scale=0.005]{Figs/icons/Python_icon_black_and_white.png}}\\Python \end{tabular} &
  
  \begin{tabular}[c]{@{}c@{}}{\includegraphics[scale=0.07]{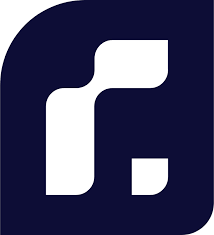}} \\ Rigetti\end{tabular} \\ \hline

\rowcolor[HTML]{FFFFFF} 
\begin{tabular}[c]{@{}c@{}}{\includegraphics[scale=0.11]{Figs/icons/1.jpg}}\\   {[}S7{]}\end{tabular} &
  \begin{tabular}[c]{@{}c@{}}{\includegraphics[scale=0.07]{Figs/icons/numb.png}}   \\ Integer Factorisation\end{tabular} &
  
  \multicolumn{1}{c|}{\cellcolor[HTML]{FFFFFF}\begin{tabular}[c]{@{}c@{}}{\includegraphics[scale=0.15]{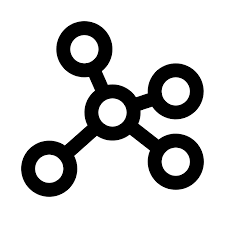}}\\Directed Graph\end{tabular}} &
  
  \begin{tabular}[c]{@{}c@{}}{\includegraphics[scale=0.04]{Figs/icons/API.png}}\\ API Gateway\end{tabular} &
  
  \multicolumn{1}{c|}{\cellcolor[HTML]{FFFFFF}\begin{tabular}[c]{@{}c@{}}{\includegraphics[scale=0.04]{Figs/icons/oPTIMISATIOPN.png}} \\Optimisation \end{tabular}} &
  
  \begin{tabular}[c]{@{}c@{}}{\includegraphics[scale=0.07]{Figs/icons/Noevidence.png}}\\No Evidence\end{tabular} &
  
  \begin{tabular}[c]{@{}c@{}}{\includegraphics[scale=0.20]{Figs/icons/amazon.png}}\end{tabular} \\ \hline

\rowcolor[HTML]{FFFFFF} 
\begin{tabular}[c]{@{}c@{}}{\includegraphics[scale=0.04]{Figs/icons/opinion.png}}   \\ {[}S8{]}\end{tabular} &

  \begin{tabular}[c]{@{}c@{}}{\includegraphics[scale=0.03]{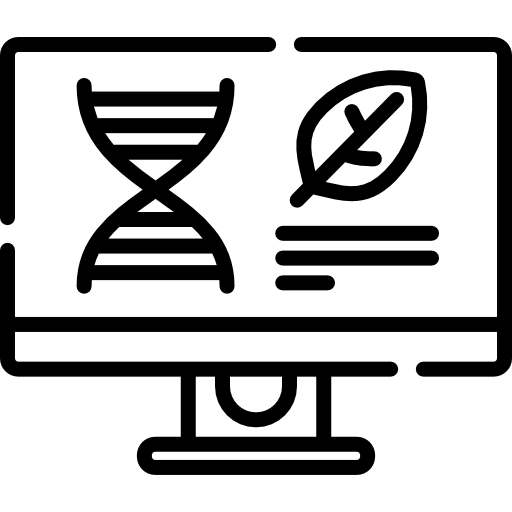}}\\Bio-inspired\\
Computing\end{tabular} &
  
  \multicolumn{1}{c|}{\cellcolor[HTML]{FFFFFF}\begin{tabular}[c]{@{}c@{}}{\includegraphics[scale=0.07]{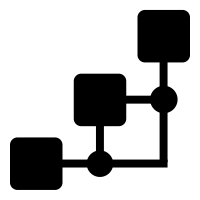}}\\Ontologies\end{tabular}} &
  
  \begin{tabular}[c]{@{}c@{}}{\includegraphics[scale=0.09]{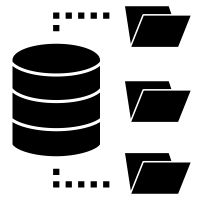}}\\Repository \\
Pattern\end{tabular} &
  
  \multicolumn{1}{c|}{\cellcolor[HTML]{FFFFFF}\begin{tabular}[c]{@{}c@{}}{\includegraphics[scale=0.04]{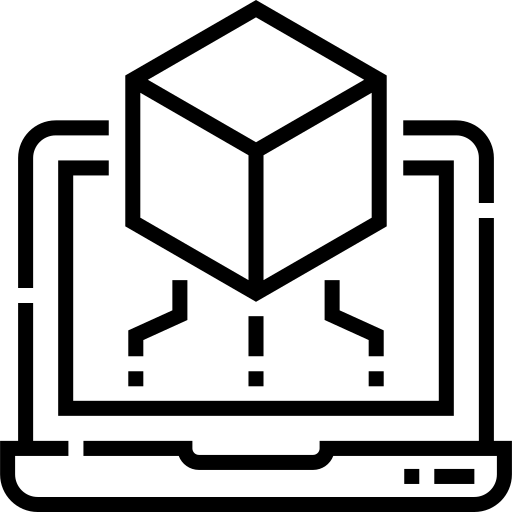}} \\Simulation\end{tabular}} &
  
  \begin{tabular}[c]{@{}c@{}}{\includegraphics[scale=0.07]{Figs/icons/Noevidence.png}}\\No Evidence\end{tabular} &
  
  \begin{tabular}[c]{@{}c@{}}{\includegraphics[scale=0.07]{Figs/icons/Noevidence.png}}\\No Evidence\end{tabular} \\ \hline

\rowcolor[HTML]{FFFFFF} 
\begin{tabular}[c]{@{}c@{}}{\includegraphics[scale=0.04]{Figs/icons/opinion.png}}   \\ {[}S9{]}\end{tabular} &
 
  \begin{tabular}[c]{@{}c@{}}{\includegraphics[scale=0.11]{Figs/icons/numb.png}} \\Integer \\Factorisation \end{tabular} &
  
  \multicolumn{1}{c|}{\cellcolor[HTML]{FFFFFF}\begin{tabular}[c]{@{}c@{}}{\includegraphics[scale=0.07]{Figs/icons/Noevidence.png}}\\No Evidence\end{tabular}} &
  
  \begin{tabular}[c]{@{}c@{}}{\includegraphics[scale=0.09]{Figs/icons/servicewrapper.png}} \\Service  
Wrapping\end{tabular} &
  
  \multicolumn{1}{c|}{\cellcolor[HTML]{FFFFFF}\begin{tabular}[c]{@{}c@{}}{\includegraphics[scale=0.07]{Figs/icons/Noevidence.png}}\\No Evidence\end{tabular}} &
  
  \begin{tabular}[c]{@{}c@{}}{\includegraphics[scale=0.005]{Figs/icons/Python_icon_black_and_white.png}}\\Python\end{tabular} &
  \begin{tabular}[c]{@{}c@{}}{\includegraphics[scale=0.20]{Figs/icons/amazon.png}}\end{tabular} \\ \hline
\end{tabular}}
\par\end{centering}
\label{tab:SMSResults}
\end{table*}

\end{landscape}

\subsection{Step III - Implementing the Reference Architecture}\label{sec:3_3}
The third step of the research method corresponds to implementing a QCaaS solution based on the reference architecture (Step II), as in Figure \ref{fig3:method}.  We refer to the implementation as developing a prototype solution, representing a QCaaS proof-of-the-concept based on the architecture.

A use case of prime factorisation is included for the implementation that exemplifies how each of the four phases and six activities adopted from Table \ref{tab:SMSResults} are utilised in the reference architecture as in Figure \ref{fig6:Implementation}. Details for reference architecture implementation are presented in Section \ref{sec:implement}.  

\subsection{Step IV - Evaluating the Reference Architecture}\label{sec:3_4}
Finally, the last step of the research method involves evaluating the reference architecture based on feedback from quantum software engineering practitioners. As per the ISO/IEC 25010 model for evaluating the quality of software-intensive systems artifacts, we evaluated the functional suitability and usability of reference architecture \cite{20_ISOQuality}. We engaged a total of 22 QSE practitioners and collected their feedback based on 17-point criteria for architecture evaluation. Details of the architecture evaluation are presented in Section \ref{sec:ArchEvaluation}.

\section{Creating the Reference Architecture for QCaaS}
\label{sec:RefArch}

Reference architectures provide a point of reference to structure a system by identifying and representing architectural components and connectors, applying reuse knowledge and best practices, and facilitating communications between domain professionals (a.k.a. system stakeholders) to design concrete architectures \cite{38_ArchEval}. We present a layered view of the reference architecture in terms of three distinct layers namely \textit{service development}, \textit{service deployment}, and \textit{service split} layers, illustrated in Figure \ref{fig5:RefArch} and detailed below. Figure \ref{fig5:RefArch} a) shows an abstract view to conceptualise the structural composition of the architecture, whereas Figure \ref{fig5:RefArch} b) uses the abstract view to present a fine-grained and concrete view of the architecture in terms of i) architectural layers, ii) phases and activities (from Table \ref{tab:SMSResults}) encapsulated inside each layer, iii) human roles, and iv) service artifacts.  A fine-granular representation of the reference architecture based on individual layers and elements encapsulated inside each layer is detailed below based on the illustrations in Figure \ref{fig5:RefArch}.   

\subsection{Architecture Layers}\label{sec:layers}
Architectural layering allows a separation of functional concerns which means architectural elements that support similar functionality can be structured in the same layer. For example, Figure \ref{fig5:RefArch} shows that conception, modelling, and assembly of quantum services can be unified into a layer named service development layer, service execution/hosting is presented as part of deployment layer, and the split between quantum and classic execution is presented as part of the service split layer.  In addition to supporting the structuring of the system, a reference architecture provides a template that helps to design a (software) solution for a particular domain, quantum computing in this case. Reference architectures also provides a common architecting vocabulary such as functional requirements and their representation as architectural components and connectors. For example, the service development layer has one of the phases named \textit{Quantum Service Conception} that encapsulates two activities namely \textit{Functional Specifications} and \textit{Quality Attributes}. These two activities allow a human role in the architecture such as \textit{Service Developer} who can specify the required functionality and desired quality of the service. The phase produces a \textit{Quantum Significant Requirements} (QSRs) as an artifact that provides the foundation for service design. In short, architectural structuring as layers contains phases that encapsulate multiple activities, enabling a human role to produce the service artifact for an incremental conception, modelling, assembly, and deployment of QCaaS.

\subsection{Phases and Activities}\label{sec:phaseactivity}
The phases and activities summarised in Table \ref{tab:SMSResults} are adopted from the SMS \cite{15_QAAS} and organised based on IBM service-oriented architecture (SOA) lifecycle \cite{IBMSOA} to develop the reference architecture. As per the SOA lifecycle for service engineering and development, each phase tells \textit{what needs to be done?} while an activity or collection of activities demonstrates \textit{how it is to be done?} For example, in Table \ref{tab:SMSResults}, the phase referred to as modeling helps to focus on how to model or represent the quantum services for their implementation, while the activities named (i) \textbf{modeling notation} show class and component diagrams as UML-based modelling notation to enable service modelling and (ii) \textbf{patterns} indicate best practices to model the QC services.

To architect QCaaS, we divided the \textbf{Model} activity from SOA life cycle into two activities namely \textit{Conception} and \textit{Modeling} to distinguish between functional needs (conception) and representation (modeling) of quantum service design. The distinction allows a fine-grained representation of the reference architecture that delineates (i) the conception of functional needs and (ii) the model that architecturally represents the functional needs for their implementation. Model represents the conception as the design specification of functional needs for quantum services. For example, the initial row of Table \ref{tab:SMSResults} highlights that the conception, i.e., required functionality that enables the delivery of quantum software is modeled, i.e., architecturally represented using a UML deployment diagram and applying the API Gateway pattern. The reference architecture in Figure \ref{fig5:RefArch} do not have the \textbf{Manage} phase from SOA life cycle as we could not find any evidence in the literature that supports identity, compliance, and business metrics management of quantum services. The phases, their underlying activities, and the data in Table \ref{tab:SMSResults} is used to create the reference architecture shown in Figure \ref{fig5:RefArch}.

\subsection{Human Roles}\label{sec:humanroles}
Human roles represent human knowledge, expertise, or activities as part of the architecting that enables development or utilisation of the quantum services \cite{31_QSASLR}. There are two types of human roles in the reference architecture, referred to as service developers and service users. Service users can be individual(s) or a group of users (part of a team or an organisation) that requires quantum computation. In contrast, service developers can include a multitude of roles that include but are not limited to quantum service developers, quantum algorithm designers, and quantum domain engineers. A recent study on architecting quantum software highlights the need for quantum-specific expertise such as quantum software architects who can map the operations of QuBits to architectural components, and quantum code managers who can simulate and analyse the flow of quantum information processing \cite{18_QSA}. A specific human role such as quantum domain engineer can analyse quantum-specific attributes like mapping between QuGates and their corresponding QuBit representation \cite{17_QSE}. Quantum domain engineer focuses on design and fabrication activities to improve QPU performance by optimising quantum algorithms and programming languages. Quantum domain engineers, guide quantum hardware and software teams to realise quantum significant requirements that need to be implemented as quantum algorithms for execution on quantum computing platforms, as in Figure \ref{fig5:RefArch} b).

\subsection{Service Artifacts}\label{sec:artifacts}
\begin{figure*}
\centering
\includegraphics[width=1.05\textwidth]{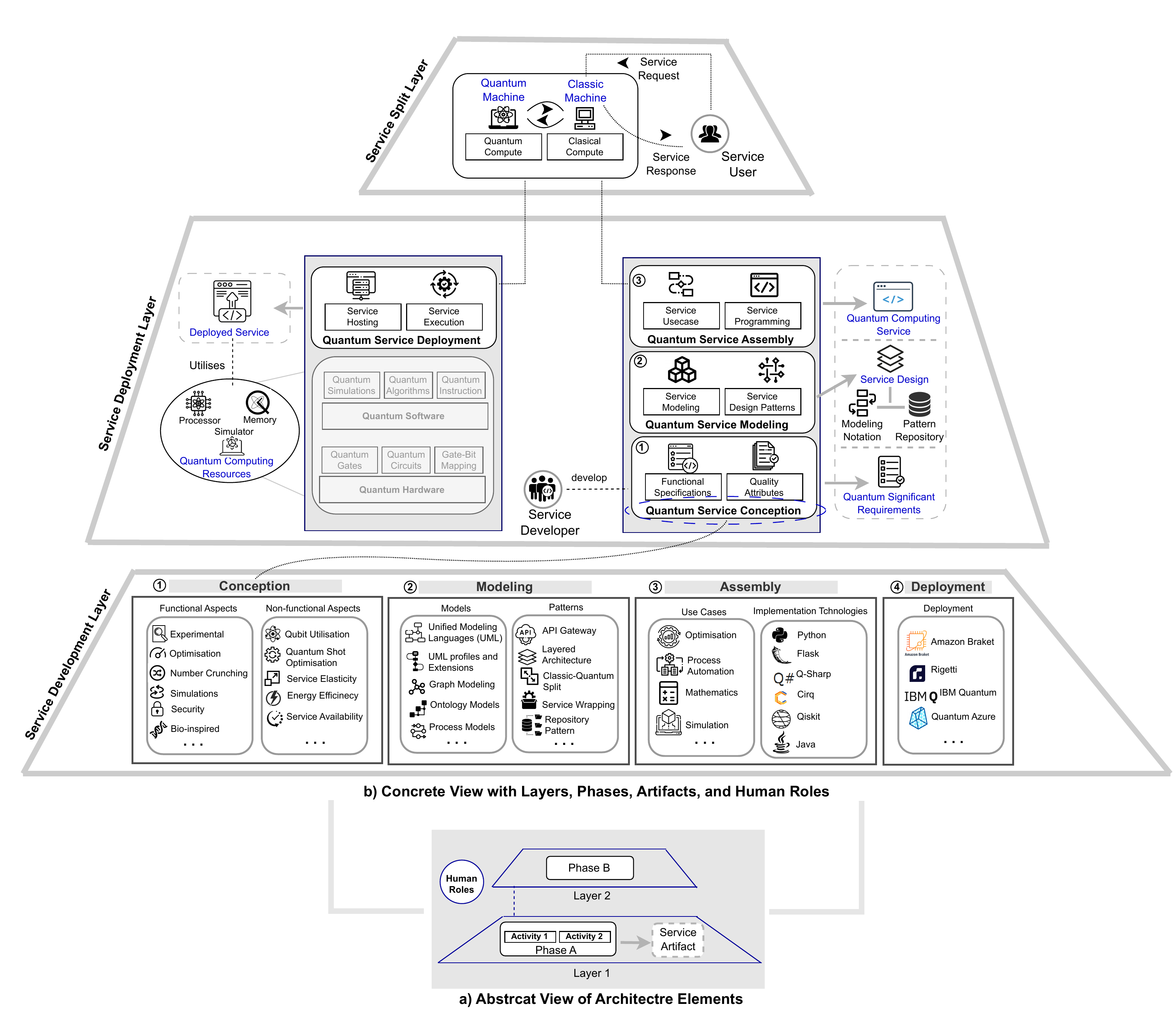}
\caption{\label{fig5:RefArch} Overview of the Reference Architecture for QCaaS.}
\end{figure*}

Service artifacts refer to any (tangible) outcome as a document, design model, or source code script that enable the development and delivery of the quantum service. As in Figure \ref{fig5:RefArch}, each phase in the architecture layer produces an artifact as an outcome of the specific phase. There are four artifacts referred to as Quantum Significant Requirements, Quantum Service Design, Quantum Service Implementation and Quantum Service Deployment. For example, the quantum service conception phase produces QSR having function and quality attributes as an artifact that supports modelling and pattern-based design of quantum service.

\section{Implementating the Reference Architecture}
\label{sec:implement}

Reference architecture-based implementation of QCaaS is highlighted in Figure \ref{fig6:Implementation} that is incremental and driven by four phases of the service lifecycle from Figure \ref{fig5:RefArch} that is based on the evidence from published research documented in the SMS and its core findings synopsised in  Table \ref{tab:SMSResults}. The implementation does not refer to a comprehensive or full-scale development of the solution, it provides a proof-of-the-concept in terms of executable specifications for quantum computing as a service. The proof-of-the-concept represents a use-case of the reference architecture that provides the basis for further implementation and validation of the architecture. The use case as a specific instance of implementation is elaborated later in this section while discussing the service assembly.


\subsection{Classical vs Quantum Microservices} \label{sec:classicMS} Microservices help software designers and developers to structure software applications, via the application of microservice architecture style, as loosely coupled and deployable modules of code that enable computation and data storage \cite{16_QAASGateway}. Microservices systems enable enterprises to adopt the service computing model by migrating or modernising their monoliths to a single application that consists of multiple small services. These microservices operate independently, with each service running in its own process and communicating, i.e., enabling message passing among each other, using lightweight mechanisms like HTTP resource APIs \cite{IBMSOA}. Industry-leading service providers such as Netflix and Amazon represent typical examples of how microservices architecture helps businesses to deliver their core business functionality (e.g., video streaming, online shopping) with increased scalability that efficiently serves millions of customers across the globe. Quantum microservicing is a term that refers to microservices that are deployed and executed on quantum computing platforms. For example, the solution \cite{34_QFAAS} utilises Qiskit an open-source quantum software development kit to develop and deliver quantum computation as a collection of microservices that can be executed on IBM Quantum platform. From an implementation perspective, which primarily involves writing and executing source code, both classical and quantum microservice are similar because they implement an algorithm or contain a module of source code that can be executed on QCs \cite{15_QAAS}. However, from operational and deployment point of view, which focuses on managing the execution of service on quantum platforms, QSRs need to be fulfilled as implemented functionality and desired quality of the microservice. Microservices that fulfill QSRs are referred to as quantum-enabled microservices or simply quantum microservices. For example, QSRs may require quantum-classic split of a quantum algorithm, efficient utilisation of QuBits, and quantum error minimisation in quantum task execution. For example, the study \cite{16_QAASGateway} suggests that a quantum algorithm can be wrapped in a microservice using the service wrapping pattern that allows classical microservices to be executed on a QC platform. 

\subsection{Service Conception}\label{sec:conception}
Figure \ref{fig6:Implementation} provides conception of quantum services by outlining the functional aspects and quality attributes, which collectively refer to as the QSRs, as specified below. The listings below highlight the functional specification of the quantum services to generate the prime factors of a given integer. To provide this functionality, the computations need to be split between a classical machine and a quantum machine. There is also a need to assess the correctness of the implementation and efficiency of the solution for prime factorisation. The listings below as the functional aspects and quality attributes two quality attributes concerning an efficient utilisation of the QC resources in terms of the utilisation of the QuBits. The quality attributes complement the functional specification with desired quality of the solution. 

\begin{mdframed}
Develop a solution that acquires an input integer N and outputs its prime factors F. 

\vspace{0.2em}
\textsf{- Split} classical and quantum computations.

\vspace{0.2em}
\textsf{- Validate} correctness and efficiency.
\end{mdframed}

\begin{mdframed}
\textsf{QuBit utilisation} The solution should efficiently utilize the available qubits by minimizing the number of qubits required for factoring integers of a given size.
\end{mdframed}





For example, Figure \ref{fig6:Implementation} indicates two architecting activities as part of service conception to specify the functional and quality aspects of prime factorisation. Functional and quality aspects are specified as quantum significant requirements that represents an architectural artifact of the conception phase. The service architect/developers can rely on the QSRs to utilise available modelling notations (e.g., UML, ADL) and apply any patterns (e.g., service orchestrator) to create a service design that acts as an artifact for assembling the microservices. 

\subsection{Service Modeling} \label{sec:modeling}
The QSRs as part of the conception act as the foundation to create a service design model that is reflective of the functional aspects. The quantum service modeling phase relies on two activities (i) creating the model, i.e., visual representation of the service and (ii) applying the pattern, i.e., the design decision to model the service. 

\textit{Modeling the quantum service} As per Figure \ref{fig6:Implementation}, we have used the UML as the modelling notation to create the service model based on UML component diagram and UML sequence diagram. To create quantum-enabled models for QSE, the quantum UML profile extends the traditional UML diagrams to support structural and behavioral modeling of quantum software \cite{32_QUML}.  Specifically, the UML component diagram in Figure \ref{fig6:Implementation} shows a structural view of the quantum services and their interconnection. Following the notations from UML profile \cite{32_QUML}, from a service modeling perspective each service is represented as an individual component. The component provides a computational service (e.g., generate a random number) and communicates with another component (s) using a connector. For example, the service named \textsf{GetGCD} interconnects with another service named  \textsf{Controller} to generate the Greatest Common Divisor (GCD) of a randomly generated number as part of the algorithmic implementation. In contrast to a structural view of the quantum service model using a component diagram, the UML sequence diagram reflects the behavior of the services in terms of message passing among the services to enable service communication.  For example, in the sequence diagram, the controller service \textsf{C: Controller} passes a message \textsf{Generate(N)} to the number generator service \textsf{N: NumGenerator} that generates a random number \textsf{R} and returns it to the controller service. A number of other UML diagrams can also be used such as the UML use case diagram to represent the QSRs. For demonstration purposes, we have used the UML component and sequence diagrams to exemplify service modeling in terms of structural composition and behavioral representation of the services to be assembled.

\vspace{0.3 em}
\textit{Pattern-based Modeling:}  Service design and development patterns provide reuse knowledge and best practices to engineer service-oriented solutions \cite{IBMSOA}. Patterns can be particularly useful, as concentrated wisdom of experienced software designers and developers, that can guide novice engineers (e.g., quantum algorithm designers, quantum code developers) to rely on existing knowledge and practices to develop quantum software effectively and efficiently \cite{25_QPattern}. In Figure \ref{fig6:Implementation}, we have applied two patterns namely the \textit{orchestrator} and \textit{classic-quantum split} pattern. The orchestrator pattern is a classic SOA pattern that helps to orchestrate the execution of a number of services to complete a service-driven task \cite{42_Orchestrator}. For example, the application  of the orchestrator pattern helped to orchestrate the factorisation of a randomly generated prime number via \textsf{NumGenerator} and \textsf{Factorise} services. The classic-quantum split pattern helps to split the functionality between quantum and classic computing, also referred to as hybrid quantum computing \cite{43_Qunatum-Classic} .  In the context of Figure \ref{fig6:Implementation}, the \textsf{Controller} service orchestrates the functionality between  classical and quantum microservices. For example, the services named \textsf{NumGenerator} and \textsf{GetGCD} can be executed on a classical machine, whereas the services \textsf{QunatumModularExponentiation}, \textsf{QunatumInverseQFT}, and \textsf{Factorise} are executed on the quantum machine. As part of the reference architecture, the modeling phase exploits modeling notation and applies a pattern to create a service design, i.e., a 'visual model' that can be assembled into executable quantum microservices.

\subsection{Service Assembly}\label{sec:assembly}
\begin{landscape}
\begin{figure}
\centering
\includegraphics[width=1.6\textwidth]{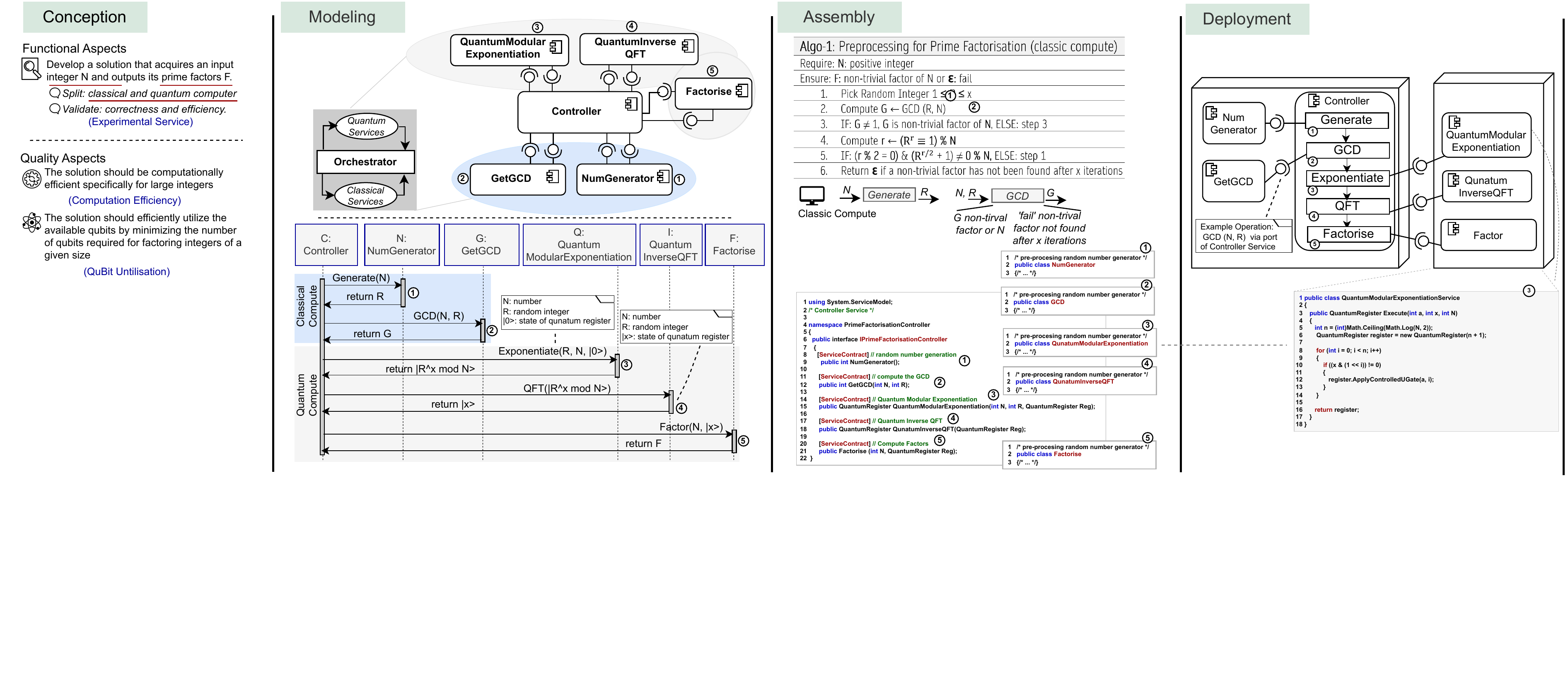}
\caption{\label{fig6:Implementation} Architecture Based Implementation of the QCaaS.}
\end{figure}
\end{landscape}

Service assembly refers to assembling a service, i.e., identifying a use-case and programming the service that implements the usecase as shown in Figure \ref{fig6:Implementation} and detailed below. 

\vspace{0.3em}
\textit{Service Usecase:} Implementation details are elaborated based on Figure \ref{fig6:Implementation} that demonstrates the conception, modelling, assembly, and deployment of software services that implement the Shor’s. Shor’s algorithm is a quantum computing algorithm that is used for factoring integers in a polynomial time \cite{26_Shor}.  Shor’s algorithm has a number of use cases such as prime factorisation, cryptography (e.g., breaking the RSA scheme), and finding the period of a function. In the scope of this work, we only focus on developing microservices, executed on a QC platform that implement Shor’s algorithm for prime factorisation. We provide an illustrative case in Figure \ref{fig6:Implementation} that exemplifies the architecture-based development of the services (from Figure \ref{fig5:RefArch}) to implement Shor’s algorithm. Specifically, Figure \ref{fig5:RefArch} as a reference architecture highlights \textit{what needs to be done?} by providing a blue-print to architect quantum service-orientation. Figure \ref{fig6:Implementation} implements the reference architecture to highlight \textit{how it is to be done?} with an illustrative example of implementing quantum service-orientation. 

\textit{Service Programming:}
In order to implement the use-case service needs to be programmed for its execution. This means that design specifications from the modeling phase need to be translated into executable specifications as part of service assembly. Figure \ref{fig6:Implementation}, shows the algorithmic design and its implementation as part of service assembly. The algorithm shows a partial assembly of services by exemplifying the classical services that generate a random number and compute the GCD for quantum computation of the Shor's algorithm. The UML sequence diagram is used as a model to assemble the service via an algorithm and its implementation in a given programming language. A synoptic view of the code snippet (written in C\#) is shown that highlights the source code skeleton for a controller service \textsf{PrimeFactorisationController} that orchestrates the classical and quantum services. 

\subsection{Deployment}\label{sec:deployment}

As the last phase in the reference architecture, deployment consists of two main activities namely service hosting and service execution. While execution and hosting primarily depend on the QC platform on which the quantum services are deployed.

\vspace{0.3em}
\textit{Service Hosting:} The hosting of the assembled service is presented as a UML deployment diagram as in Figure \ref{fig6:Implementation}. UML deployment diagram show two deployment nodes (hosting machines). The classical compute node hosts three services \textsf{NumGenerator}, \textsf{GetGCD}, and \textsf{Controller}, whereas the quantum compute node hosts three services namely \textsf{QunatumModularExponentiation}, \textsf{QunatumInverseQFT}, and \textsf{Factor}.

\begin{figure}[h!]
\centering
\includegraphics[width=0.5\textwidth]{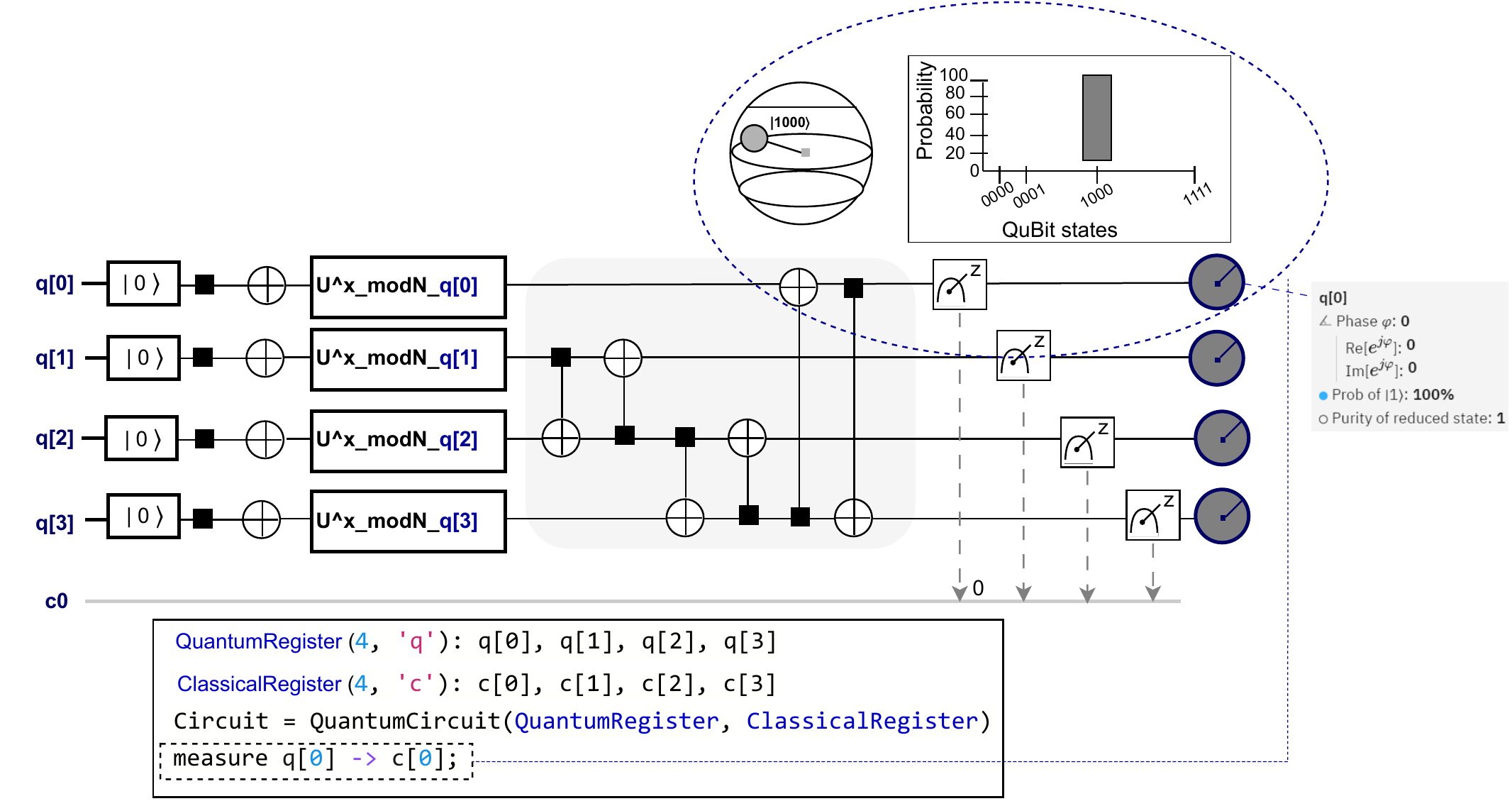}
\caption{\label{fig7:Circuit} Circuit Diagram for QuBit Utilisation.}
\end{figure}

\vspace{0.3em}
\textit{Service Execution} depends on the platform on which the quantum service has been deployed. We have used the Miscrosoft Qunatum Azure platform to deploy the qunatum microservices for prime factorisation. Figure \ref{fig7:Circuit} shows the circuit diagram for utilisation of the QuBits for quantum service execution. The notation \textsf{q[0] \ldots q[3]} shows the allocation of Qubits and \textsf{c[0] \ldots c[3]} represent classical bits. All the Qubits are initiliased such as $U\textsuperscript{$\wedge$} x\_modN\_q[0]$ with a unitary operator for modular exponentiation acting on the first QuBit. The measurement results of each qubit are stored in the classical bits \textsf{c[0] \ldots c[3]}. These classical bits are then used for classical post-processing, such as applying the continued fraction algorithm to extract the factors. This is expressed as: Circuit = QuantumCircuit(QuantumRegister, ClassicalRegister) to measure q[0] -> c[0].

\section{Practationers' Evaluation of the Reference Architecture}
\label{sec:ArchEvaluation}
This section details evaluation of the reference architecture based on practitioners’ perspectives to assess the suitability and usability of the architecture (per ISO-IEC-25010 quality model \cite{20_ISOQuality} in the context of software services for QCs. First, we highlight the architecture evaluation methods and distinguish between the evaluation of a concrete and a reference architecture in Section \ref{sec:evalmethods}. We then introduce the practitioners who evaluated the architecture in terms of their geo-diversity, professional roles, years of experience, domain of experience etc. regarding quantum service computing in Section \ref{sec:demography}. The results of architecture evaluation, driven by a structured questionnaire and practitioners’ feedback are detailed in Section \ref{sec:surveyanswers}.
\vspace{0.3 em}
\subsection{Evaluation Methods and Concrete vs Reference Architectures} \label{sec:evalmethods}
\textit{Architecture evaluation methods:} Academic research and industry use-cases highlight two primary approaches to evaluate the reference architecture based on (i) scenario-based evaluation using architecture evaluation method \cite{39_ArchEvals} or (ii) evaluation by the domain experts using evaluation workshops or surveys \cite{40_ConcreteRef}. Evaluation methods such as Software Architecture Analysis Method (SAAM) or Architecture Trade-off Analysis Methods (ATAM) rely on scenarios that represent use-cases for evaluating the architecture. For example, the SAAM method can represent an architecturally significant requirement as an evaluation scenario by expressing: \textit{`\ldots generate a random integer  that must be stored and processed to its factorisation (i.e., functionality: `number generation' and 'factorisation')}. Architecture evaluation should validate this scenario by assessing if the random number has been generated and is factored. Evaluation method like SAAM helps architects to develop scenarios, prioritise them, and assess the impacts of scenarios as part of the architectural evaluation. Similarly, methods like ATAM rely on architectural scenarios to evaluate the trade-offs (e.g., usability vs efficiency) of the system under development. In contrast to the evaluation methods, evaluation surveys or workshops focus on presenting the architecture to domain experts such as software architects or developers in a particular domain (e.g., pervasive computing, mission-critical software) to review and evaluate the architecture based on pre-defined criteria such as fit for purpose or functional completeness. The evaluation criteria can be formulated as a collection of survey questions to be answered by the practitioners. For example, the studies provide a checklist \cite{41_EvalCheckList} and empirically derived methods \cite{38_ArchEval} that allow the practitioners (system stakeholders) to evaluate the reference architecture for embedded software and e-contracting systems. We followed the empirical guidelines from \cite{39_ArchEvals} and considered the distinction between architectural abstractions (concrete vs reference) \cite{40_ConcreteRef} to evaluate the reference architecture.

\begin{figure*}
\centering
\includegraphics[width=0.80\textwidth]{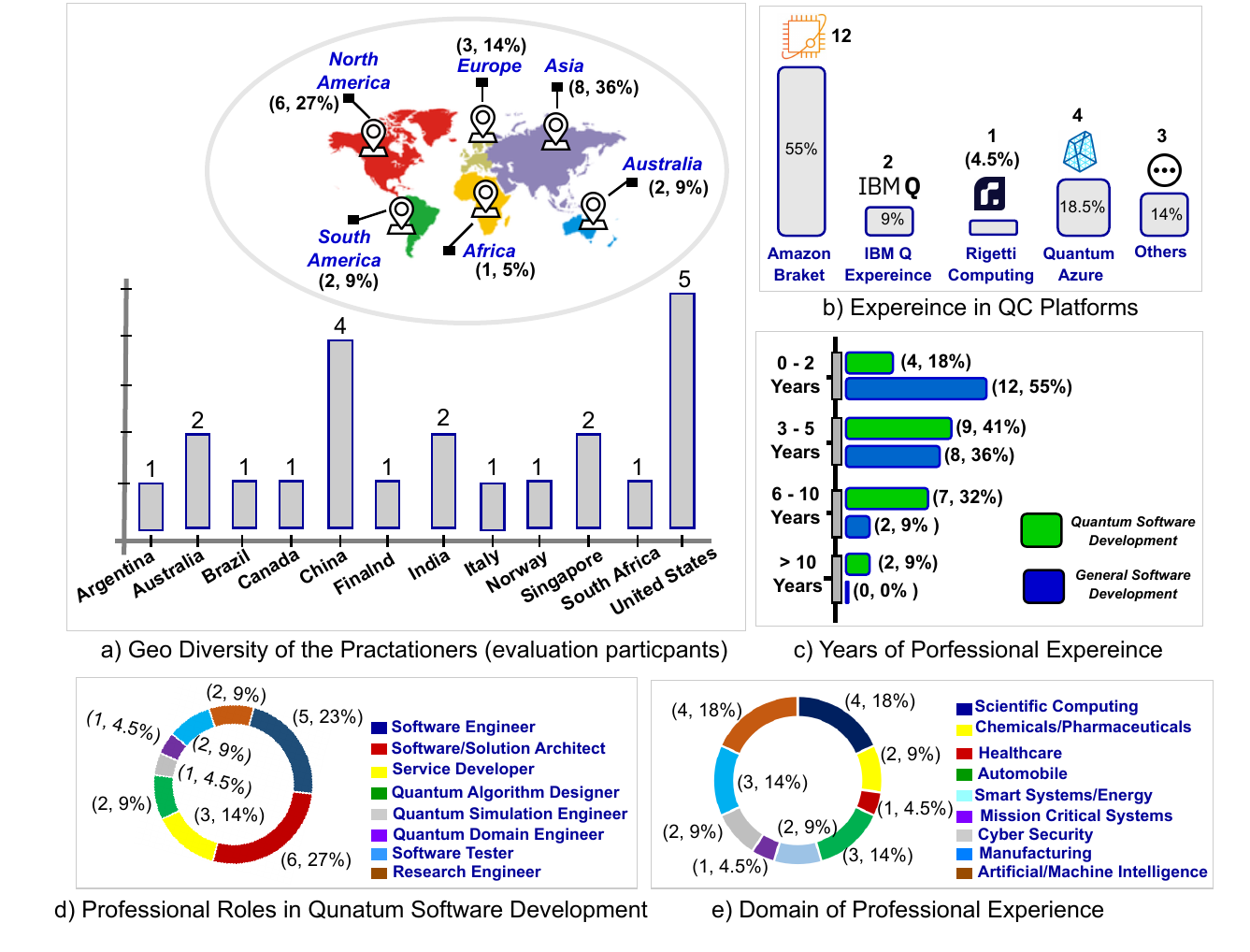}
\caption{\label{fig5:Demography} Overview of the Demographic Details of Practitioners.}
\end{figure*}

\textit{Evaluating concrete vs reference architectures:} From a software evaluation perspective, the study \cite{40_ConcreteRef} differentiates between two types of architectures namely concrete architecture and reference architecture. A concrete architecture, as exemplified in Figure \ref{fig6:Implementation} (component and connector view) is derived from a reference architecture in Figure \ref{fig5:RefArch}. The concrete architecture expresses the structural composition, functionality, and constraints if any of the software to be implemented as per the architectural description. In comparison, the reference architecture provides a generic template or a point of reference that provides architectural vocabulary such as modelling notations, layers, patterns etc. to express the architecture or simply instantiating a concrete architecture. Due to a generic nature of them, reference architectures provide an abstraction of concrete architecture and existing architecture evaluation methods such as SAAM or ATAM etc. needs to be altered to evaluate reference architectures. Research in \cite{38_ArchEval, 39_ArchEvals} shows that evaluation methods need to be extended or altered by accommodating evaluation criteria such as completeness, applicability, and buildability etc. from stakeholders’ point of view. For emergent classes of software such as blockchains or quantum software there may be a lack of evaluation methods and their underlying scenarios. In case of a lack of evaluation methods, practitioners survey is one possibility to evaluate the architecture. By adhering to ISO-IEC-25010, we used functional suitability and usability as the evaluation criteria for the reference architecture. ISO-IEC-25010 standard aims to determine the quality characteristics of software artifacts, systems, and products \cite{20_ISOQuality}, further detailed in Section \ref{sec:surveyanswers}.

\subsection{Demography Details of the Practitioners} \label{sec:demography}
Demography details of the practitioners, who evaluated the reference architecture, are visualised in Figure \ref{fig5:Demography} which highlights five aspects relating to geo-/professional diversity of the participants each detailed below. Each individual practitioner (P) was assigned a unique identifier referred to as (P1, P2, $\ldots$, P22). While collecting data based on practitioners' feedback, the idea of saturation can provide some practical guidance for estimating sample sizes, prior to data collection, necessary for conducting qualitative evaluation. As per the guidelines in \cite{44_Saturation}, conducting 12 to 15 interviews of a homogeneous group is adequate to reach saturation of data sampling. Demography details in Figure \ref{fig5:Demography} complement the evaluation and provide a fine-granular interpretation of the evaluation results. For example, P12 identified as a practitioner with an experience of less than 2 years as a quantum algorithm designer, has worked with Amazon Braket (QC platform), to develop mission-critical systems.  P12 indicated that \faCommentDots[regular] \textit{`$\ldots$ quantum algorithm for simulation or optimisation can be wrapped inside microservices and it helps to translate quantum software development challenges as a problem of developing microservices $\ldots$'} and \faCommentDots[regular] \textit{`$\ldots$ based on my prior experience with AWS (Amazon Web Services), Amazon Braket is the preferred quantum computing provider as it provides a comprehensive platform to build, test, deploy amazon web services for quantum computers $\ldots$'}

\begin{itemize}
    \item \textbf{Geo-distribution} indicates the diversity of practitioners in terms of their geo-location, as in Figure \ref{fig5:Demography} a). Geo-distribution indicates feedback from practitioners across the globe, reflecting diverse participation in architecture evaluation. Geo-distribution and diversity of QC professionals can be vital in an attempt to minimise the challenge of qunatum divide \cite{10_QDivide}. A total of 22 practitioners evaluated the architecture, participating from 12 different countries across  6 continents. For example, the QC practitioners from the United States (5/22, 23\% approx.) and China (4/22, 18\% approx.) combinedly represented a total of 41\% of all the participants of the architecture evaluation. 
    
    \item \textbf{Usage of QC platforms} indicates practitioners’ experience working with different QC platforms or quantum service providers, also referred to as the quantum vendors. As per Figure \ref{fig5:Demography} b), practitioners indicated their experience with four different types of QC platforms  with an overwhelming majority (i.e., 12/22, 55\%) have experience with Amazon Braket, a finding that is consistent with the results of our mapping study on QCaaS \cite{15_QAAS}. The mapping study highlighted that Amazon Braket (a managed Amazon Web Services (AWS)) is the most preferred platform to design, test, and run quantum algorithms. One of the reasons for selecting Amazon Braket by practitioners for service deployment is that it can allow service users/developers to design their own quantum algorithms. This can be particularly handy for novice developers unfamiliar with the technicalities of quantum systems to utilise a set of pre-built algorithms, tools, and documents to develop and manage quantum services on Amazon platform. Two practitioners, i.e., 14\% approx. indicated ‘Others’ as the QC platforms of their experience.
    
    \item \textbf{Years of experience} highlights practitioners' professional experience to engineer (i) general class of software such as web or mobile systems, and (ii) quantum software and services, as in Figure \ref{fig5:Demography} c). The professional experience quantified as the number of years highlights that a total of 12/22, i.e., 55\% of the participants have an experience of two years or less. On the contrary, 73\% of practitioners’ reflected experience between 3 to 10 years with the development of general software systems.  
    
    \item \textbf{Professional roles} indicate practitioners’ expertise in engineering and development of software systems and/or services for quantum software as in Figure \ref{fig5:Demography} d). Practitioners’ responses identified a total of 8 roles highlighted in Figure \ref{fig5:Demography} d) including but not limited to software/solution architect, quantum algorithm designer, and quantum domain engineer. Quantum domain engineer is considered as a QSE-specific role that enables software engineers to map the requirements of domain with software systems such as manipulating QuBits expressed as the configuration of the software components and services \cite{18_QSA}. Software engineers, software/solution architects, and service developers were identified as the predominant roles in quantum service development, collectively representing 14/22, i.e., 64\% of all the participants.  
    
    \item \textbf{Domain of experience} refers to the context or area of application that requires quantum software to perform computation or enable automation in a particular domain, highlighted in Figure \ref{fig5:Demography} e). For example, automobile indicates a domain for which quantum software or services can be developed to provide simulation, optimisation, or decision support for researching and developing automobile technologies. Scientific computing, artificial intelligence/machine learning, and automobile represent the predominant domain of professional expertise indicated by a total of 11/22, i.e., 50\% of the participants.
\end{itemize}

Demography details summarised above can enrich the analysis of architecture evaluation results. Specifically, demography data as extended details can help us understand if practitioners’ perspectives on architecture evaluation may be influenced by certain factors such as their years of experience, professional roles, or knowledge of any specific QC platforms. For example, one of the practitioner expressed \faCommentDots[regular] \textit{` $\ldots$ my experience with microservices development, in particular working with (Amazon) Braket that is effectively an Amazon managed web service endorses layering a quantum system into classic and quantum parts. Layering or split in this case may be intuitive, however; such a split could be counter-productive if ill-designed. What I mean is that theoretically, it looks appropriate to structure your system into the classic and quantum parts, split of a quantum program (pre-processing) and then merge the compute results (post-processing) may be time-consuming and it may result into an anti-pattern'.}

\subsection{Results of the Practationers' Evaluation} \label{sec:surveyanswers}

\begin{figure*}
\centering
\includegraphics[width=1.00\textwidth]{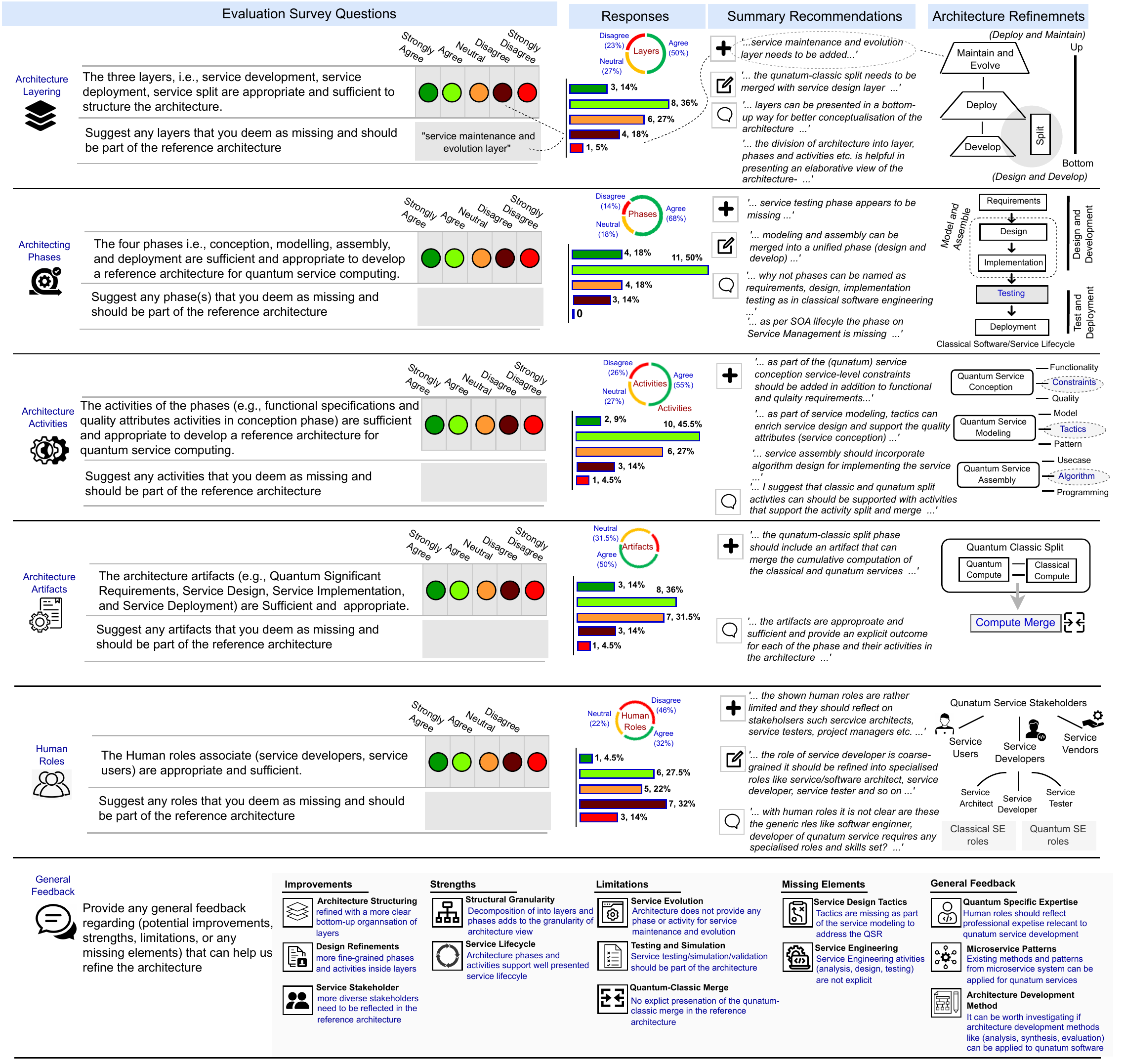}
\caption{\label{fig6:EvalResults} Summary of Results Based on Practationers' Evaluation.}
\end{figure*}

The results of architecture evaluation based on the practitioners' feedback and its analysis are summarised in Figure \ref{fig6:EvalResults}. For the clarity of presentation, the results, as per Figure \ref{fig6:EvalResults} are organised along two dimensions, i.e., survey questions (presented at the vertical axis) and practitioners' responses to the questions (presented at the horizontal axis), detailed below.

\subsubsection{Presenting the Survey Questionaire} \label{subsec:questions} The survey questionnaire comprising of a total of 11 questions (Q1 - Q11) to evaluate architecture in terms of architectural elements including architecture layering, architecting phases, architecture activities, architecture artifacts, human roles, and general feedback, presented on the horizontal axis. For example, as shown in Figure \ref{fig6:EvalResults}, the question on architecture layering seeks the practitioner's view if: \textit{`the three layers, i.e., service development, service deployment, service split are appropriate and sufficient to structure the architecture'}. The question (Q1) follows up by seeking any suggestion (Q2) by asking: \textit{`Suggest any layers that you deem as missing or should be part of the reference architecture'}. One of the survey questions (Q11) aims to gather spontaneous feedback on the overall architecture by asking: \textit{`Provide any general feedback regarding (potential improvements, strengths, limitations, or any missing element) that can help us refine the architecture'}.

\subsubsection{Presenting the Survey Questionaire} \label{subsec:responses} 
The practitioner's feedback to the questionnaire is visually summarised along the horizontal axis that highlights 4 types of information. The information includes a five-scale Likert option for each question (range: \textsf{Strongly Agree} to \textsf{Strongly Disagree}), bar and pie graphs to quantify the Likert response, a summary of the recommendations provided by the practitioners, and a view of potential refinement that could be made to the architecture based on the recommendation. For example, in response to Q1, i.e., is the presenting layering sufficient and appropriate to architect quantum service-orientation, the practitioners responded as: Strongly Agree (3 respondents: 14\% approx.), Agree (8, 36\%), Neutral (6, 27\%), Disagree (4, 18\%), and Strongly Disagree (1, 5\%). This reflects a cumulative agreement to sufficiency and appropriateness of the layering as 50\%, neutrality as 27\% and 23\% disagreement. As an example, one of the suggestions corresponding to Q2 about the architecture layering was about the addition of \faCommentDots \textit{`... service maintenance and evolution layer'} that appears as missing as per a response from one of the practitioners. The recommendations (Q2, Q4, Q6, Q8, Q10) complementary to each of the Likert question (Q1, Q3 etc.) generally represent any addition, removal, modification or general improvement comments. The suggestion about addition, removal, or modification of any architecture elements are viewed as modification operations. For example, the suggestion \faCommentDots \textit{`... service maintenance and evolution layer needs to be added...'} is classified as an addition of a specific architecture element, an additional layer in this case. Similarly, the suggestion \faCommentDots \textit{`...the quantum-classic split layer needs to be merged with service design layer...'} represents a modification of the existing architecture. The general suggestion \faCommentDots \textit{`...layers can be presented in a [classical] bottom-up way for a better conceptualisation of the architecture ...’}.

Q11 represents an exception as it seeks a relatively spontaneous response, as overall suggestions for improvements, identified strengths, limitations, any missing elements, or any conclusive general feedback about the reference architecture. We exemplify some of the responses. In terms of any suggested improvements, one of the practitioners suggested \textsf{Architecture Structuring} can be \faCommentDots[regular] \textit{`refined with a more clear bottom-up organisation of the layers'}. \textbf{Service Lifecycle} presentation reflects a strength of the architecture where \faCommentDots[regular] \textit{`architecture phases and activities support well-presented service lifecycle'.} One of the identified limitations was missing \textbf{Testing and Simulation} suggesting \faCommentDots[regular] \textit{`service testing/simulation/validation should be part of the architecture'}. One of the missing factors relates to \textbf{Service Design Tactics} \faCommentDots[regular] \textit{`tactics are missing as part of the service modelling to address the QSR'}. Finally, the General Feedback highlighted a multitude of suggestions such as investigating and integrating \textbf{Architecture Development Method} indicating \faCommentDots[regular] \textit{`it can be worth investigating if architecture development methods like (analysis, synthesis, evaluation) can be applied to quantum software'}.

\section{Related Work}
\label{sec:RelatedWork}
We now discuss the related work in terms of published research on architecting quantum software (Section \ref{sec:QSA}) and quantum service computing (Section \ref{sec:QSC}). From the QSE perspective, a synoptical view of the most relevant research helps us to contextualises the scope and contributions of the proposed research in terms of architecting quantum computing as a service. 

\subsection{Architecting Software-intensive Systems for QCs}\label{sec:QSA}
Quantum software engineering entails a multitude of engineering activities that can range from quantum domain modelling or quantum algorithmic design to quantum simulation management to support a systematic and incremental development of quantum software and services \cite{2_QSE, 17_QSE}. QSE acts an umbrella to support different phases of software development that include but are not limited to architecting \cite{18_QSA}, programming \cite{22_QPL}, and testing \cite{31_QST} of quantum software. The role of quantum software architectures in QSE becomes pivotal as it allows designers and developers to map the QSRs to a software model that is independent of technical implementations and acts as a blue-print to implement and validate quantum software. A recently conducted systematic review of quantum software architectures indicates the role that architectural processes, patterns, and tools play to empower the role of software developers to design and implement quantum software \cite{31_QSASLR}. The systematic review indicates that QSE is a relatively new engineering paradigm and often software engineers and developers find themselves underprepared and often lack the expertise to tackle quantum-specific software development challenges. In-line with the findings of the systematic review in \cite{31_QSASLR}, the research in \cite{18_QSA} presents an architectural process composed of a number of architecting activities and highlights the needs for professional expertise (i.e., human roles in QSA) to effectively develop quantum software. To support architecture-centric engineering and development, Quantum UML profile as an extension of the Unified Modeling Language (UML) supports modelling and architecting quantum software \cite{32_QUML}. Quantum UML-based models (e.g., class, use-case, deployment diagrams) can be transformed into implementation or executable specification using quantum model-driven engineering \cite{19_MDQSE}. A well-crafted architecture can abstract the implementation details via architectural components and can assist developers to achieve model-to-code transformations by exploiting model-driven QSE \cite{32_QUML, 18_QSA, 19_MDQSE}. Considering architecture as a point of reference for QSE, there is no reference architecture that can consolidate system design knowledge in terms of patterns, modelling notations, service use case, and human-roles to support architecture-centric design and development of quantum service computing.

\subsection{Quantum Service Computing}\label{sec:QSC}
A recently conducted mapping study on quantum computing as a service investigated existing research and identified some trends for future research that unifies quantum computing and service-oriented software \cite{15_QAAS}. Results of the study indicate that emerging and future work on quantum service computing relies on patterns and tactics, low-code development, and agile practices for quantum service design and implementation. The status-quo on quantum service-orientation reflects experimental research on implementing quantum algorithms \cite{33_QuantAlgo} and functions \cite{34_QFAAS} implemented as microservices that can be executed on QC platforms such as Amazon Braket. Although the use cases for quantum service computing (e.g., number crunching, mathematical optimisation) are rather limited, however, work is in progress to exploit QC platforms in software services context of bio-inspired computing \cite{35_QHumanities}. Specifically, the study proposes a tool-chain for quantum cloud computing where computation-intensive tasks can be outsourced from classical machines to quantum servers that are configured via quantum cloud computing model \cite{30_IBMCloud}. Some studies have demonstrated that existing development processes (e.g., DevOps) and service architectural styles (microservices) can be extended and successfully applied to quantum service computing \cite{36_AgileQSE}. Moreover, a number of classical patterns such as service wrapping or API gateway have been successfully applied to design software-intensive systems and services for quantum computing \cite{16_QAASGateway}. These classical patterns when combined with quantum specific patterns such as quantum-classic split can help develop a catalogue of patterns that can help novice developers and engineers to rely on reuse knowledge and best practices for quantum service-orientation driven by microservices. Quantum service computing is being envisioned as a utility computing model that can breach the quantum divide by offering QC resources such as QPU, memory or simulators to end-users. However, there is a lack of processes and patterns to enable a systemaised development of quantum service-oriented solutions. Recent studies on quantum service systems indicate the challenge of empirically investigating the extent to which classical service-orientation knowledge can be reused in the context of QCaaS \cite{15_QAAS, 16_QAASGateway, 34_QFAAS, 33_QuantAlgo}. The reference architecture can enable engineers to abstract complexities of quantum source code into architectural components, apply reuse knowledge via quantum software patterns, and adopt best practices such as microservices architecture style to develop QCaaS. Proof-of-the-concept as a prototype can enable QC users to discover and utilise hardware, software, and networking resources offered by QC vendors via quantum software services \cite{IBMSOA}. 

\vspace{0.3 em}

\textit{Conclusive Summary:}
Based on the review of the most relevant existing research, we conclude that architectural models help software engineers to tackle design and development issues by abstracting complex and implementation specific (i.e., source coding details) with high-level architectural components \cite{31_QSASLR}. Research and development on QSE \cite{2_QSE} in general and QSA \cite{18_QSA} to be more specific highlights that lack of professional expertise and unique challenges of quantum software development require developers to rely on reference models, patterns, and processes to develop quantum software services. The proposed solution complements the emerging research on QSE and QSA and more specifically focuses on an empirically grounded reference architecture that acts as a point of reference or a blueprint to implement quantum service orientation. The proposed solution draws inspiration from recommendations and guidelines from a generic model in \cite{35_QHumanities} for quantum cloud computing and pattern-based service development \cite{25_QPattern} to architect and implement quantum software services. The proposed solution advocates the need to establish reference architectures that can provide foundations to apply architectural knowledge, process, and principle that can synergise classical and quantum approaches for quantum service development. 
\section{Threats to the Validity of Research}
\label{sec:Threats}
This study draws empirically-grounded evidence from SMS and uses practitioners' feedback for evaluation thus inheriting some threats to the validity \cite{37_Threats}. These threats represent potential limitations, constraints, or flaws in the study that can impact various aspects like the generalisation, replicability, and validity of results. Any future research that relies on the presented study in terms of research design or its results must consider these threats. Future work should focus on minimising these threats to ensure methodological rigor and generalisation for avoiding any bias in the results. We discuss three main types of threats, each detailed below.

\subsection{Threate I - Internal Validity}\label{sec:threat1}
It examines the extent to which any systematic error (bias) is present in the design, conduct, and analysis etc. of the study. In the context of internal validity, we refer to the research method (Figure \ref{fig3:method}) that overview different steps to design and conduct the study to present its results. To minimise this threat, we derived the reference architecture from a mapping study \cite{15_QAAS} that followed guidelines of evidence-based software engineering to objectively collect and analyse the data \cite{27_Mapping}. Based on the collected data, we aligned the evidence from the mapping study to the phases and activities of IBM SOA lifecyle to create the reference architecture for quantum service-orientation \cite{IBMSOA}. We relied on an architecture use case to develop a proof-of-the-concept implementation of the reference architecture. The steps as part of research methods aim to minimise the bias and threats to internal validity. However, more work is required to understand and assess if the study results can be validated with a different architecting process or by adopting other evaluation methods.

\subsection{Threate II - External Validity}\label{sec:threat2}
It examines whether the findings of a study can be generalised to other contexts. From an external validity perspective, it needs to be determined if the same process can be applied to develop other reference architectures or this reference architecture can be used to develop other systems in a quantum computing context. We only experimented with a single case study of prime factorisation and only two patterns  classic quantum split and orchestrator patterns of moderate complexity that can compromise the study's generalisation. Specifically, scenarios with the increased complexity of architecting process (quantum simulations), types of patterns (microservice patterns), and human expertise (novice/experienced engineers) can affect the external validity of this research. We did try to minimise the external validity by engaging 22 QSE practitioners and their feedback to improve the suitability and usability of the reference architecture \cite{20_ISOQuality}. Future work requires more rigorous evaluation, preferably in a more practical industrial context to further assess the external validity of the research.

\subsection{Threate III - Conclusion Validity}\label{sec:threat3}
It determines the degree to which the conclusions reached by the study are credible or believable. In order to minimise this threat, we followed a three-step process (Figure \ref{fig3:method}) to support a fine-grained process to architect (Figure \ref{fig5:RefArch}) the software and validate the results (Figure  \ref{fig6:Implementation}). Moreover, a case study-based approach was adopted to ensure scenario-based demonstration of the study results. However, some conclusions (e.g., practitioners’ perspective, practical context) can only be validated with more experimentation involving multiple case studies, and real context scenarios of architecting quantum service computing.
\section{Conclusions}
\label{sec:Conclusions}
Quantum computing has started to emerge as a disruptive technology – striving to offer quantum computational supremacy over traditional digital computers - by exploiting hardware, software, and networking technologies that are driven by the operations of quantum bits and quantum gates. QCs are in a phase of their inception, however; they have started to demonstrate their computing superiority in areas that range from bio-inspired systems, data security and cryptography solutions to tackling optimization problems. A plethora of issues such as hardware limitations, lack of software ecosystems, scarcity of human expertise to engineer QCs ecosystem, and quantum error rate impede a wide-scale adoption and commercially viable solutions of quantum computing. This research aims to synergise quantum software engineering and service computing to architect quantum service orientation for pay-per-shot usability of QC resources. Specifically, the research focuses on unifying quantum software engineering methods and service-orientation patterns to promote reuse knowledge and best practices to tackle emerging and futuristic challenges of architecting and implementing Quantum Computing as a Service (QCaaS).   

\textit{Primary contributions and implications:} The primary contribution of this research include (a) an empirically-derived reference architecture as a blue-print to develop software services for QCaaS, (b) a proof-of-the-concept that demonstrates architecture-centric implementation of quantum software services, and (iii) practitioners’ evaluation of the reference architecture that provides recommendations and guidelines to design and develop solutions for QCaaS. The research can have implications for researchers and practitioners of quantum software engineering. Specifically, the results of the study help academic researchers to understand the role of reference architectures and quantum service-orientation the challenges of QSE in the context of QCaaS. Moreover, the practitioners can explore the reference architecture as a system blueprint and patterns as reuse knowledge that can be adopted to develop solutions for QCaaS.

\textit{Needs for future research:}  Based on the study results, we envision future work in two directions with a focus on empirical research including (a) mining social coding platforms and (b) practitioners' interview to further understand the architecting and development of quantum computing as a service. Specifically, by mining social coding platforms (e.g., GitHub) we can empirically discover knowledge and understand the practices adopted by developers’ communities in open-source QCaaS. The study provides foundations to design and conduct semi-structured interviews by engaging service developers and engineers to seek their feedback and synthesise the results as practitioners’ perspectives to complement the evidence from design and implementation of the reference architecture.

\bibliographystyle{ieeetr}
\bibliography{References}

\end{document}